\documentclass[aps,pre,reprint,superscriptaddress,amsmath,showpacs]{revtex4-1}

\usepackage{graphicx}
\usepackage{xcolor}
\usepackage[utf8]{inputenc}
\usepackage[T1]{fontenc}
\usepackage{placeins}
\usepackage{hyperref}
\usepackage{booktabs}

\newcommand{\bl}{\color{blue}}

\begin{document}

\title{
Supply chain network rewiring dynamics at the firm-level
}

\author{Tobias Reisch}
\affiliation{Complexity Science Hub Vienna, Vienna A-1080, Austria}

\author{András Borsos}
\affiliation{Complexity Science Hub Vienna, Vienna A-1080, Austria}
\affiliation{Department of Financial Systems Analysis, Central Bank of Hungary, Budapest 1013, Hungary}

\author{Stefan Thurner}
\email{Corresponding author: stefan.thurner@meduniwien.ac.at}
\affiliation{Complexity Science Hub Vienna, Vienna A-1080, Austria}
\affiliation{Section for Science of Complex Systems, CeMSIIS, Medical University of Vienna, A-1090 Vienna, Austria} 
\affiliation{Santa Fe Institute, Santa Fe, NM 85701, USA}
\affiliation{Supply Chain Intelligence Institute Austria, Vienna A-1080, Austria}

%


\begin{abstract}
Supply chain networks (SCN) form the structural backbone of any society. They constitute the societal metabolism that literally produces everything for everybody by coordinating practically every single person on the planet. SCNs are by no means static but undergo permanent change  through the entry and exit of firms and the re-arrangement of supply relations. Here we use a unique dataset to explore the temporal evolution of firms and their supplier-buyer relations of a national SCN. Monthly reported value added tax data from Hungary from 2014 to 2022 allows us to reconstruct the entire economy with 711,248 companies and 38,644,400 connections, covering practically every re-structuring event of an entire economy at firm-level resolution. We find that per year about 25\% of firms exit the SCN while 28\% new ones enter. On average, 55\% of all supply-links present in one year will not be present in the next. We report the half-life time of supply-links to be 13 months. New links attach super-preferentially to firms with a probability, $p(i)\propto k_i^{1.08}$, with $k_i$ firm $i$'s number of supply-connections. We calibrate a simple statistical network generation model that reproduces the stylized characteristics of the dominant Hungarian SCN. The model not only reproduces local network features such as in- and out-degree distributions, assortativity and  clustering structure, but also captures realistic systemic risk profiles. We discuss the present model in how rewiring dynamics of the economy is essential for quantifying its resilience and to estimate shock propagation. 
\end{abstract}

%
%
%
%

\flushbottom
\maketitle
\thispagestyle{empty}

The economy, i.e. the invention, production, distribution, consumption, usage, management, infrastructure, recycling and disposing of almost all intermediate and final goods and services is organized through firms. At the firm-level most decisions in the economy are taken, for example what and how to produce their goods or services, who to hire, when to invest, how to innovate and how to do administration, etc. Firms are connected to each other through buyer-supplier relations. Firms, together with the material and financial flows on the buyer-supplier links, form the backbone of every society's metabolism -- literally. They manage essential information of the flows of  goods, products, production, investments, ideas, services, payments, etc. These relations not only produce all goods, services, food, buildings, and infrastructure, they also organize, educate, and maintain talents and workers, investments, etc. 

\begin{figure*}[t!]
	\centering
	\includegraphics[width=0.99\textwidth]{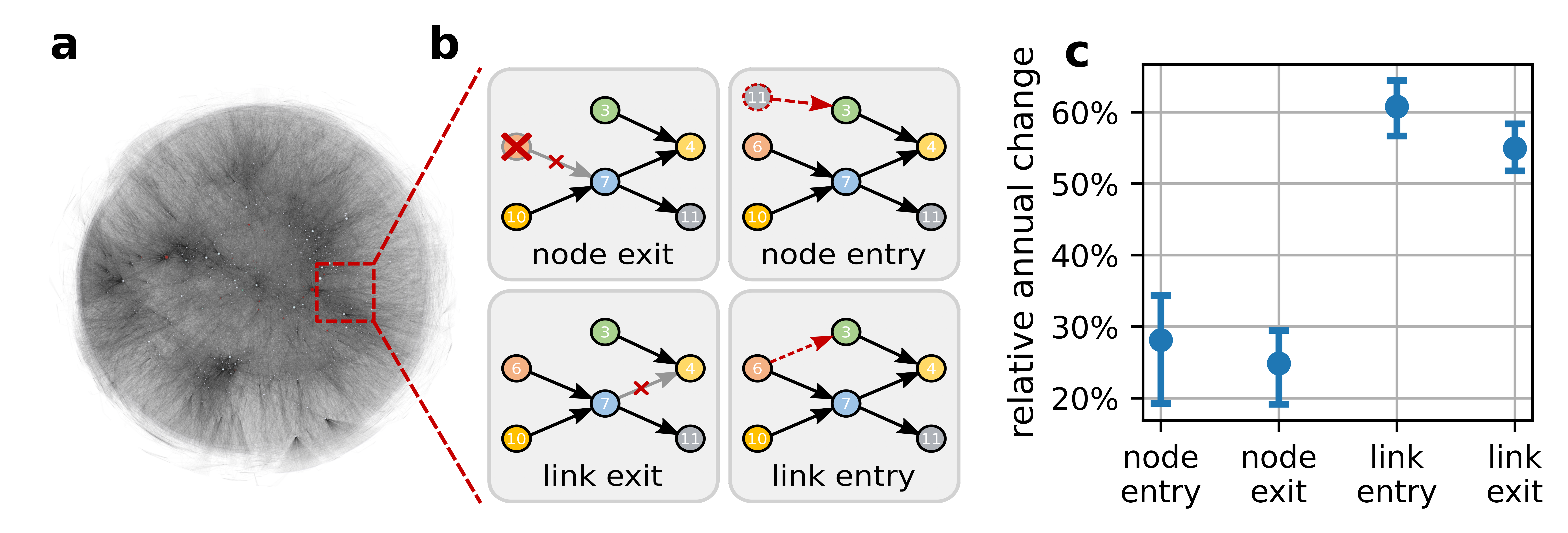}
	\includegraphics[width=0.99\textwidth]{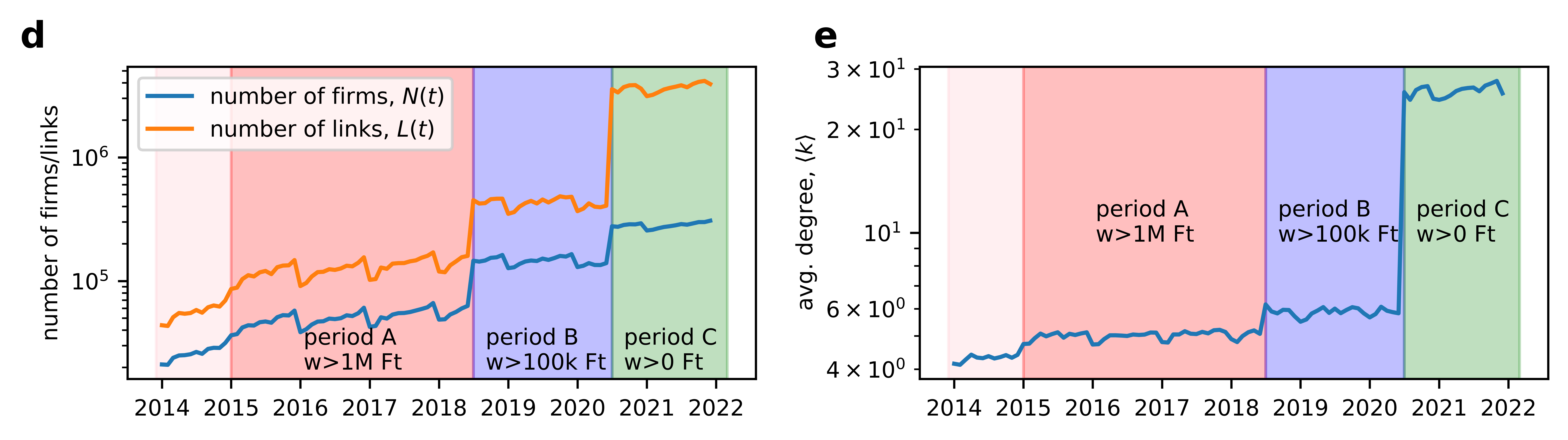}
	\caption{Temporal evolution of the Hungarian production network. 
		(a)  Snapshot of the Hungarian SCN in 2017. Nodes represent 91,595 companies connected by about 230,388 supply relations. Layout: force-and-spring.
		(b) Schematic view of the four elementary network re-configuration steps. 
		(c) Relative annual rates of the network re-structuring processes. Symbols show the average rate of change with respect to the previous year, bars denote the minimum and maximum values across the years 2015-2022, where we exclude the years 2018 and 2020, because there were changes in the reporting standards that distort the rates.
		(d) Monthly number of active firms and links. The red, blue, and green shading highlight periods with different reporting thresholds for the links.
		(e) Average degree for each month. Same shading as (d).
		Note that despite the massive numbers of nodes and links entering and exiting, the network size and link density remain stable within periods with no change in reporting threshold.}
	\label{fig:concept_figure}
\end{figure*}

The set of all buyer-supplier relationships within an economy are often referred to as \textit{supply chains}, which is to some extent misleading, since most production processes (or sequences of labour steps) are not structured as simple linear chains, but these `chains'  intersect, and constantly change over time to form complex structures that we call \textit{supply chain networks} (SCN). For a depiction of a temporal snapshot of a national SCN, see Fig. \ref{fig:concept_figure}a. SCN are central for understanding many economic processes such as innovation~\cite{ernst2002global, schilling2007interfirm}, growth~\cite{mcnerney2022production}, development~\cite{hidalgo2007product}, greenhouse gas emissions~\cite{wiedmann2009review, davis2010consumption, stangl2025}, economic shock spreading~\cite{barrot2016input,carvalho2021supply,diem2022quantifying}, and resilience~\cite{henriet2012firm, klimek2019quantifying, inoue2019firm}. Surprisingly little is known about the economy at this `atomistic' scale of SCNs, in particular the structures, patterns, and laws behind the dynamical rewiring of the associated networks are hitherto unknown.  

Economies are comprised of several hundred  thousands to millions of companies that are connected through hundreds of millions to billions of buyer-supplier dependencies. Globally, there are an estimated 300 million firms with an estimated 13 billion supply links~\cite{pichler2023building}. Until recently, it seemed unimaginable to investigate the SCNs of entire economies at the firm-level. The standard today is to look at an aggregation of companies into so-called industry sectors, such as the NACE industry classification scheme \cite{naceclassification2006}. First attempts to map a national economy on the industry level date back to Leontief's work almost a century ago~\cite{leontief1941structure}, which lead to the framework of \textit{input-output economics} that still today is a workhorse for applied economics. Only recently empirical descriptions of SCNs on the firm-level arrived. The first studies, based on a large commercial dataset from Japan~\cite{saito2007larger,ohnishi2009hubs,fujiwara2010large}, report structural network features such as a scale free degree distribution, disassortativity, and sub linear scaling of sales with the number of supply connections.
An early description of a European national SCN uses value added tax (VAT) data of Belgium to understand the individual firms' network distance to final demand and their relationship with international trade~\cite{dhyne2015belgian}. In recent years, several large datasets, many based on VAT data, were accumulated, including Hungary~\cite{borsos2020unfolding,diem2022quantifying}, Ecuador~\cite{bacilieri2023firm}, or Uganda~\cite{spray2017reorganize}; for a recent review, see~\cite{bacilieri2023firm}. In countries where such data is not collected, SCNs have been reconstructed from alternative data sources, such as statistical surveys~\cite{hooijmaaijers2019methodology,mattsson2021functional}, monetary transactions~\cite{ialongo2022reconstructing}, or inter-firm communication networks~\cite{reisch2022monitoring}. For a review of the dramatic growth of datasets in recent years in terms of firms involved, see \cite{pichler2023building}.

Most studies consider the SCNs as directed, weighted networks~\cite{saito2007larger,ohnishi2009hubs,fujiwara2010large,dhyne2015belgian,borsos2020unfolding,bacilieri2023firm}, while only some recognize their multi-layer nature ~\cite{borsos2020unfolding,diem2022quantifying,hooijmaaijers2019methodology,mattsson2021functional,ialongo2022reconstructing} that associates different products, services, or industries with layers of a multi-layer network. The multi-layer structure of SCNs is used as a key ingredient for reconstruction tasks from alternative data sources~\cite{hooijmaaijers2019methodology,ialongo2022reconstructing,reisch2022monitoring}.

To arrive at an atomistic understanding of the economy working at a systemic level, knowing the structure of national SCNs is not sufficient. It is necessary to know how firms turn their inputs into  outputs, i.e., their \textit{production functions}. Production functions can be of different type, depending on the nature of the firm: it can be of Leontief type, meaning that the output is determined by the minimum of the (essential) inputs weighted by the so-called technical coefficients \cite{diem2022quantifying}. Another type of production function is called linear, where the output is a linear combination of inputs. The removal of one input does not stop the output entirely. In \cite{diem2022quantifying} a functional combination of Leontief and linear production functions was introduced; we will employ this more general production function in the remainder of the work. The maybe most popular production function used in economics
is the Cobb-Douglas production function \cite{carvalho2019production}, a power law, i.e., a product of inputs taken to (input-specific) powers. 

An essential observation is that when combining the SCN topology with the production functions of the constituent firms the network is turned into a \textit{hypergraph}, a generalized network structure that connects sets of nodes (inputs) with other sets (outputs) \cite{thurner2018introduction}. The hypergraph structure is especially important for understanding the operation of SCNs on the systemic level, including  shock propagation in production networks. It is essential for the appropriate quantification of \textit{economic systemic risk}~\cite{diem2022quantifying}. In the following we use the term `SCN' and keep its hypergraph structure in mind.

The final step for an atomistic understanding of the economy is to understand the dynamics of its SCNs and production functions. SCNs are subject to continuous restructuring, or rewiring. In Fig. \ref{fig:concept_figure}b we show the four elementary processes that take place when a SCN evolves in time. Firms can exit and enter the production network at certain rates. Whenever a firm exits all of its buyer-supplier connections (in- and out-links) vanish; if a new firm enters it establishes new links to existing firms. Buyer-supplier links typically are constantly updated between firms (link exit and link entry) both, in terms of strength (amount of goods/services exchanged), and who trades with whom. The underlying attachment mechanisms behind these highly dynamical link-updates are typically non-trivial and include details like firm strategy and technological decisions~\cite{oberfield2018theory}, price differences~\cite{gualdi2016emergence}, geographic proximity of firms~\cite{bernard2019production}, current network structure~\cite{chaney2014network,carvalho2014input}, personal preferences and taste of decision makers and are therefore hard to quantify and might ––at the current state–– only be accessible on statistical grounds. 

Here we explore the temporal evolution of an empirical SCN on statistical grounds, by examining every entry and every exit of firms as well as the formation and termination of every buyer-supplier connection during almost a decade of the economic history of Hungary from 2014 to 2022, see Materials \& Methods. To this end we use monthly reported VAT data in Hungary containing 711,248 companies and 38,644,400 connections. We cover practically every re-structuring event  of an economy at the firm-level resolution. We quantify monthly entry and exit rates for both firms and links and study the local conditions in the SCN under which link changes occur, in particular the degree of nodes that enter. For links that are generated between firms, we estimate the role of both the industry sector, and the size of customers and suppliers. Building on these results, we then develop a simple network generative model to understand the network properties that emerge from the microscopic restructuring processes. We include various aspects of previous network generative models, including~\cite{dorogovtsev_structure_2000,moore_exact_2006,johnson_nonlinear_2009} and choose a sparse parametrization that is exclusively based on microdata. The model covers firm entry and exit, spontaneous link generation and exit, as well as non-trivial attachment processes for buyer-supplier connections, see Results section. We then compare the structural network characterisitcs of the model with the observed empirical data, such as various degree distributions, assortativity structure, and local clustering. Finally, we test whether the model is able to realistically capture systemic risk and compare model systemic risk profiles with empirical ones. The employed systemic risk measure is the Economic Systemic Risk Index (ESRI)~\cite{diem2022quantifying}, a network centrality measure specifically designed to estimate the systemic risk created by every single firm in a SCN -- up- and downward the SCN. ESRI  takes production functions explicitly into account. Our model is data-driven in the sense that we calibrate {\em all} parameters directly from microdata -- there are no free {\em ad-hoc} chosen parameters remaining in the model.

There has been a series of previous attempts to model SCN formation under various aspects. To understand the in-degree distribution of the US production network, a generative network model was developed using firm-level microdata~\cite{atalay2011network}. Another early contribution~\cite{saavedra2008asymmetric} studies the decline of the New York garment industry and finds that preserving asymmetric (disassortative) links (in terms of degree) is important to retain the topology and functionality of the shrinking network. Subsequent studies~\cite{chaney2014network,carvalho2014input}  emphasized the importance of the presently realized network structure for its evolution. A model where firms adopt new suppliers based on price differences proposes an explanation for the Zipf-type distribution of firm's out-degrees~\cite{gualdi2016emergence}. In a more theoretical exercise, a model was studied where suppliers are selected based on both the match-specific productivity and the cost of the associated input, resulting in the emergence of ``star suppliers" with a large number of customers~\cite{oberfield2018theory}. An empirical analysis of the input-output structure of the European Union using a Stochastic Actor Oriented Model (SOAM)\cite{mundt2021formation} tries to synthesize several previous studies and finds that supplier heterogeneity in productivity, growth, labor costs, and several structural network properties --such as firm degree, the tendencies to connect to indirect suppliers, and to form reciprocal relationships-- all play a role in the evolution of production networks. Earlier studies based on SOAMs were used to study the role of geography on the formation of firm ties~\cite{balland2012proximity,balland2013dynamics}. A recent model~\cite{ozaki2024integration} combines a generative network model for the network topology based on~\cite{miura2012effect} with a diffusion model for the link weights~based on \cite{tamura2018diffusion-localization}. In terms of data, previous studies are either purely theoretical~\cite{gualdi2016emergence,oberfield2018theory}, calibrated on sector-level input-output data~\cite{chaney2014network,mundt2021formation}, focused only on single sectors~\cite{saavedra2008asymmetric,balland2012proximity,balland2013dynamics}, or fitted --at least partially-- to macroscopic network properties such as the degree distribution of firm-level data~\cite{atalay2011network,carvalho2014input,ozaki2024integration,bernard2022sparse}.

\begin{figure*}[t]
	\centering
		\includegraphics[width=0.9\textwidth]{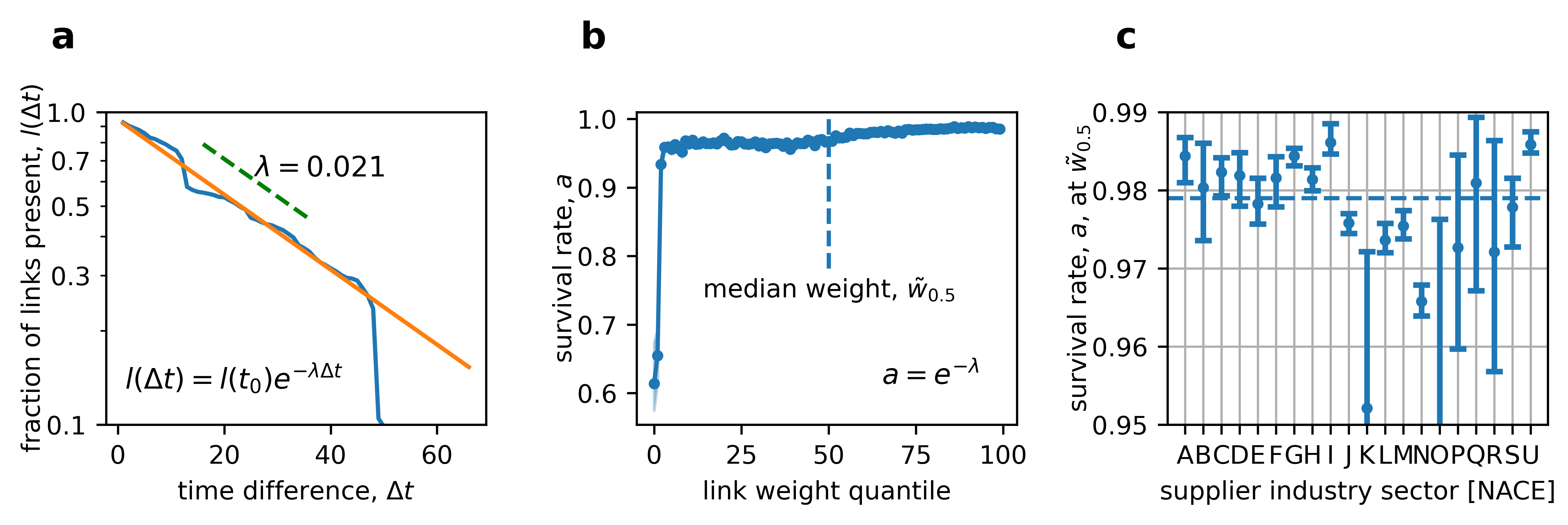}
	\caption{Estimation of link decay rates in the Hungarian SCN. 
	(a) To estimate the persistence of supply links we identify all links that exist at time $t_0 = \mathrm{Jan. }2017$ and plot the fraction, 
	$l(\Delta t)$, that is still active at time $t_0 + \Delta t$, $l(\Delta t) = L(t_0 + \Delta t)/L(t_0)$. We fit an exponential decay $l(\Delta t) = 
	L(t_0) e^{-\lambda \Delta t}$ (orange line), in semi-log scale to obtain the monthly decay rate, $\lambda=0.021$.
	(b) Survival rate, defined as $a = e^{-\lambda}$ as a function of link weight percentiles. The blue line denotes the average survival 
	rate, the shaded area is the 90\% CI.
	(c) Survival rate for the NACE sections of the supplying firm at the median link weight. The dashed, horizontal line denotes the  
	average overall survival rate of 95 \%. The link exit is well described by an exponential decay that has a lower rate for higher link 
	weights and is heterogeneous for different industries.}
     \label{fig:link_decay}
\end{figure*}

\section*{Results}

\begin{figure}[t!]
    \centering
    \includegraphics[width=0.85\columnwidth]{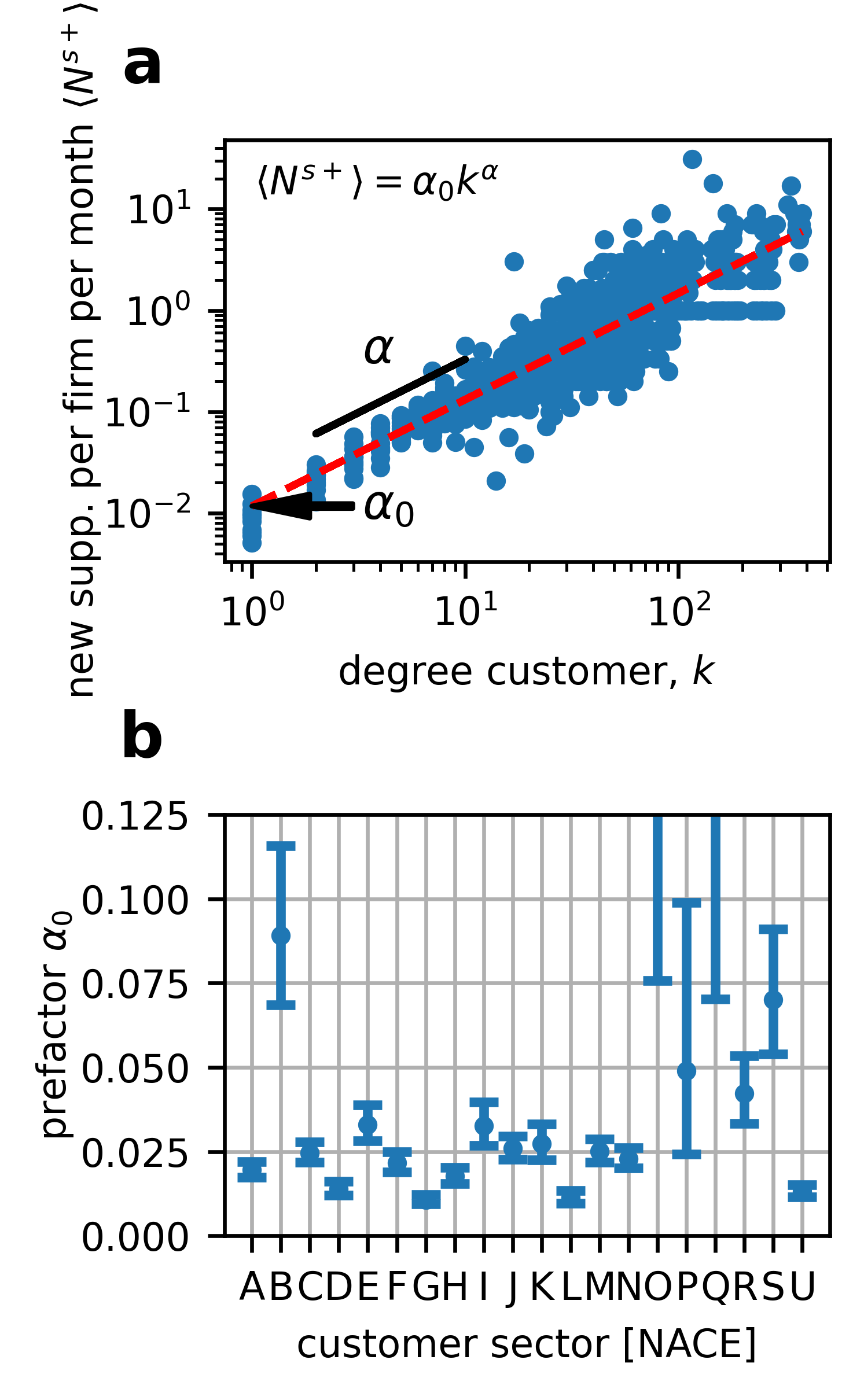}
    \caption{In-link generation overall and for industry sectors.
             (a) Average number of new suppliers per firm per time step (month) as a function of the degree, $\langle N^{s+} \rangle$, for every 
             degree and month in 2017. The red line shows the best fit line $\langle N^{s+} \rangle = \alpha_0 k^\alpha = 0.012 k^{1.049}$.
             To assess differences in supplier-turnover, we fix the scaling exponent $\alpha = 1.0$ and fit the linear slopes, $\alpha_0$, for 
             every NACE sector. Bars denote the 90\% CI. For most industries the linear slope is between 0.01. and 0.04, only for industries 
             with a very few  (below ca. 50 observations, identical with those that show large errorbars) $\alpha_0$ is larger.}
    \label{fig:spontaneous_link_generation}
\end{figure}

\subsection*{SCN characteristics and reporting thresholds}
We study the Hungarian SCN based on VAT data records from 2014 to 2022, see Materials \& Methods. Figure~\ref{fig:concept_figure}d shows the number of firms (blue), $N(t)$, and connections (orange), $L(t)$, active in each month, $t$. Both quantities show pronounced jumps from June to July in 2018 and 2020 that are caused by a successive lowering of reporting thresholds (above which a VAT payment enters the data), see Materials \& Methods. We name the periods alphabetically as indicated in Fig.~\ref{fig:concept_figure}d. The change of reporting thresholds introduces a systematic bias that we cannot correct for, and we separately analyze the three periods highlighted in Fig.~\ref{fig:concept_figure}d. We find an average $\bar{N}(A)=50,615$ and $\bar{L}(A)=127,274$ in period A, $\bar{N}(B)=129,835$ and $\bar{L}(B)=378,037$ in period B, and $\bar{N}(C)=282,879$ and $\bar{L}(C)=3,669,802.$ in period C. Sub-annually both, $N$ and $L$, show a seasonality, where the lowest numbers occur in January and the highest in December. 

In Fig.~\ref{fig:concept_figure}e we plot the average degree $k(t) = L(t)/N(t)$ for every month. With lower reporting thresholds the network becomes denser, from $\bar{k}(A)=5.0$ in period A, $\bar{k}(B)=5.7$ in period B, to $\bar{k}(A)=25.9$ in period C. The most substantial densification occurs from B to C, suggesting a large number of low-transaction-value supply connections. The relative sub-annual change is smaller than for $N$ and $L$ in panel d, suggesting that the fluctuations in those quantities are driven by node activity.

\subsection*{Firm- and link-turnover}
Figure~\ref{fig:concept_figure}c shows the average relative annual change rates for the four processes described before in Hungary. Between 2015 and 2017 (period A) we find that per year about 25 \% of firms exit the Hungarian VAT network and 28 \% new firms enter; the number of firms effectively grows with 3.3 \%. On average, 55 \% of all links present in one year are not present the next year. However, relative to the previous year, also 61 \% of new links appear, resulting in an effective growth of 5.8 \%. The average link is found to have a half-life time of 13 months. These numbers indicate a massive, ongoing restructuring of the SCN. In SI Text 1 we investigate the link turnover rates by supplier and customer industry. We find very different link entry and exit rates depending on the specific supplier-customer industry combination.

\subsection*{Estimation of entry and exit rates}
We start by analyzing the entry and exit rates for firms and links, respectively, in the {\em persistent network}; for its definition, see Materials and Methods. In Fig.~\ref{fig:entry_exit_rates} a we plot the empirical probability density function (PDF) of the monthly number of firms that enter the production network in 2017 (blue). The dashed vertical line denotes the empirical average of $348.9$ new firms per month. The solid orange line shows the PDF of a superposition of Poisson processes that takes the annual seasonality in rates properly into account, see SI Text 2. Figure~\ref{fig:entry_exit_rates}b shows the number of firm exiting per month in 2017 (blue). The dashed vertical line denotes the empirical average at $167.6$ removed firms per month. In Fig.~\ref{fig:entry_exit_rates}c and d we plot the histogram of links entering and exiting per month in 2017, respectively. The  dashed orange lines denote the averages at $867.8$ for links entering and $606.9$ for links exiting per month. All four processes, entry and exit of links and firms, can be modeled well with Poisson processes (orange lines). Here we report overall firm and link turnover rates, not distinguishing whether a link has vanished due to firm exit or otherwise. We will distinguish these cases below.

\subsection*{Firm entry}
In the model, firms enter the network according to an Poisson distribution, with the mean rate matching the observed average monthly entry of new firms. Upon entry, each firm establishes connections with both suppliers and buyers. These connections are quantified by the firm's in-degree (number of supplier links) and out-degree (number of buyer links). New firms in our dataset show an average in-degree of 0.35 and out-degree of 0.71. Notably, we observe a correlation between in-degree and out-degree, which we capture in our model by sampling from the empirical joint distribution illustrated in Fig.~\ref{fig:SI_degree_upon_entry}. The node's sector is randomly drawn with the probability of sampling a sector proportional to its prevalence in the network,
\begin{equation}
p(s) = n(s) / N \quad \text{,}
\label{eq:entry_sector_probability}
\end{equation}
where $n(s)$ denotes the number of firms in sector, $s$, and $N$ the overall number of nodes.

\subsection*{Firm exit}
The model implements firm exits in two distinct ways, calibrated to match the observed exit rates from our dataset. In the first mechanism, firms are selected for removal with a uniform probability $p^{ex}$, along with all their incoming and outgoing connections. The second mechanism removes isolated firms -- those without any connections -- at the end of each timestep. The probability, $p^{ex}$, is calibrated to match the empirical exit rate, $p^{\text{exit empirical}}$, using the relation
	\begin{equation}
	p^{\text{exit empirical}} = p^{ex} + \sum_{k} p(k) (p^{\text{remove link}})^k \quad \text{,}
	\label{eq:node_removal_prob}
	\end{equation}
where the first term denotes the first uniform node removal probability and the second the probability of a node exiting due to removal of all its links. Together, these exit mechanisms successfully reproduce the left-skewed distribution of monthly firm exits observed in the data, as shown in Fig.~\ref{fig:SI-model_DeltaNL}a.

\begin{figure*}[t]
    \centering
    \includegraphics[width=0.9\textwidth]{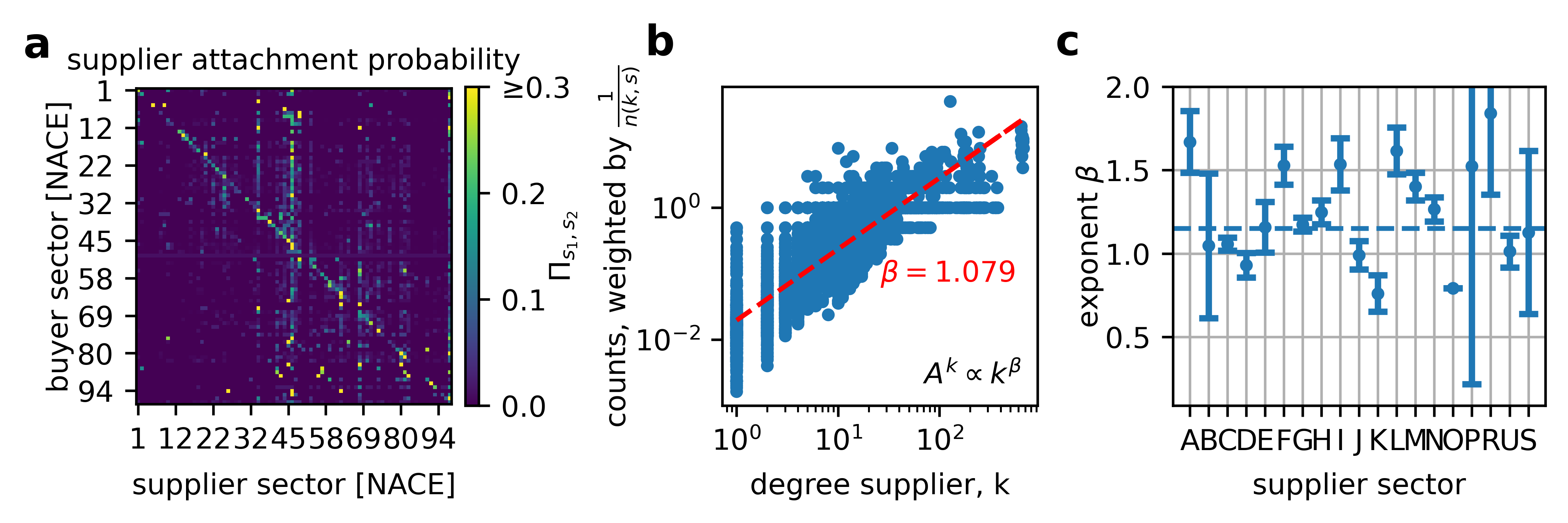}
    \caption{Estimation of the process how links attach to suppliers. We estimate the link attachment in 2 steps. We start with 
    (a) The supplier attachment probability matrix, $\Pi_{s_1  s_2}$ for 2017, showing the probability that a firm in industry $s_1$ will choose 
    a supplier from sector $s_2$. The matrix is clearly sparse, the diagonal and a few vertical lines are visible.
    (b) Estimation of the attachment kernel exponent, $\beta$. 
    The attachment kernel, $A^k(k)$, determines the probability of a customer choosing a supplier $i$ in a given sector $s_2^*$ as a   
    function of $i$'s degree, $p(i) \propto A^k(k_i)$. The number of new out-links attaching to nodes with a given degree $k^*$ is proportional 
    to the nodes' attachment kernel and the number of nodes with degree $k^*$, $n(k^*, s^*)$. We can estimate $A^k$ by plotting the 
    number of new out-links attaching to firms with with a degree $k^*$ inversely weighted by $n(k^*, s^*)$. 
    The dashed red line shows the best fit to a scaling law $A^k(k) = \beta_0 k^\beta = 0.02 k^{1.079}$. 
    (c) Scaling exponents, $\beta$, for the NACE sections, with 90\% confidence intervals. The horizontal dashed line corresponds to the 
    slope in b. The exponents show a lot of heterogeneity, ranging from 0.76 to 1.84, but are clearly larger than 0, highlighting preferential  
    attachment.
 }
    \label{fig:link_attachment}
\end{figure*}

\subsection*{Link exit}
In Fig.~\ref{fig:link_decay} we study the link exit process. To understand the persistence of supply links we identify all links at time $t_0 = \text{Jan. 2017}$, $L(t_0)$, and plot the fraction of them that is still active at time $t_0 + \Delta t$, $l(\Delta t) = L(t_0 + \Delta t)/L(t_0)$ (blue line). We fit an exponential decay $l(\Delta t) = e^{-\lambda \Delta t}$ (orange line). For $t_0 = \text{Jan. 2017}$, we find an overall (including all links) $\lambda = 0.021$. This can be transformed to the survival rate, $a = e^{-\lambda} = 0.979$ or the link removal probability of 
\begin{equation}
	p^{\mathrm{remove link}} = 1-a = 0.021
	\label{eq:link_removal_prob} \quad \text{.}
\end{equation}
This means 97.9\% of all links that exist in one month  also exist the following month and the yearly survival rate is thus $77.5$ \%. This is significantly higher than reported in Fig.~\ref{fig:concept_figure}c because we are studying the dominant network, which is more stable than the overall change considered in Fig.~\ref{fig:concept_figure}c. We check if the link exit process can be described better by a process with a changing decay rate; see SI Text 3. We find that the data is best described by the process with constant decay rate.

How the survival rate (or decay constant) depends on the link weight (transaction volume), we show in Fig.~\ref{fig:link_decay}b where the survival rate, $a$, is plotted as function of the link weight quantile, the shaded area denotes the 90\% confidence interval (CI). Below the 20th percentile of link weights $a$ is low, increasing from  0.614 for the lowest to 0.965 for the 20th percentile. Above the 20th percentile we observe a formation of a plateau  (with only a small increase) to 0.988 for the links with the highest volume. At the median link weight, $\tilde{w}_{0.5} = 188 \text{million Forint}$ (approximately 480,000 EUR), we find $a_{\tilde{w}_{0.5}}=0.968$. Clearly, links with higher volume become more stable, above the bottom quintile the survival rates become practically constant; links in the bottom quintile are much more volatile and volatile and volume dependent.

To clarify the dependence of the survival rate on the industry cluster of the supplying firm, we fit the decay rate at median link weight for every supplier industry sector independently (using section level NACE~\cite{naceclassification2006} codes) and show the results in Fig.~\ref{fig:link_decay}c (bars denote the 90\% CI). The variability of $a$ between sectors is high, with values ranging from $a_O = 0.875$ in  sector O to $a_I = 0.986$ in  sector I. See Tab.~\ref{tab:SI_NACE_letters} for the descriptions of the NACE codes.  There manufacturing sectors (B-F) tend to have higher survival rates than service sectors (G-U). For sectors with very few observations (as in K, O, P, Q and R), the CI is large.
Sectors 
``I - Accommodation and Food Service Activities",
``A - Agriculture, Forestry and Fishing", and 
``G - Wholesale and Retail Trade; Repair of Motor Vehicles and Motorcycles" are more stable than average, while sectors 
``N - Administrative and Support Service Activities"
``K - Financial and Insurance Activities", and
``O - Public Administration and Defence; Compulsory Social Security", appear to form relatively short lived connections. 
Note that for sectors N and O there are not many observations and the confidence intervals are large. In Tab.~\ref{tab:SI_decay_rates}, SI Text 3, we provide a detailed table of sector level survival rates. The situation gets slightly more involved for link entries, where one has to distinguish between suppliers and customers.

\subsection*{Link entry -- customers}
\label{sec:link_entry_cust}
When links enter the network not all firms have an equal probability to appear as customers or suppliers. We first analyze link entry from the perspective of a customer by calculating the average number of new suppliers per firm per month, $\langle N^{s+}\rangle$. Later we investigate to which firms these links attach. In Fig.~\ref{fig:spontaneous_link_generation}a we show $\langle N^{s+} \rangle$ as function of degree, $k$, and using OLS on the log-variables we fit a power function, 
\begin{equation}
	\langle N^{s+} \rangle = \alpha_0 k^{\alpha} 
	\label{eq:new_links_per_cust}
\end{equation}
and obtain an approximate scaling relation. The fit (red dashed line) yields $\alpha_0 = 0.012 \pm 0.002$ and $\alpha = 1.049\pm 0.027$, where we report the 90\% CI.

To estimate the baseline rate at which firms acquire new suppliers by sector, we fix the scaling exponent to $\alpha = 1.0$ and fit $\alpha_0$ separately for each customer industry. Fixing $\alpha = 1.0$ corresponds to assuming that (in a steady state) suppliers are replaced at the same rate by firms of all size. In Fig.~\ref{fig:spontaneous_link_generation}b we plot $\alpha_0$ by industrial sector using NACE sections, the error bars denote the 90\% CI. For sectors with a few observations the error is large and results are not useful.
Several sectors have turnover rates that are significantly larger than average, in particular (listing only sectors with small confidence intervals)
``E - Water Supply; Sewerage, Waste Management and Remediation Activities", and
``I - Accommodation and Food Service Activities", and
``K - Financial and Insurance Activities". 
Sectors with significantly slower than average supplier turnover are 
``D - Electricity, Gas, Steam and Air Conditioning Supply", 
``G - Wholesale and Retail Trade; Repair of Motor Vehicles and Motorcycles", 
``L - Real Estate Activities". In Tab.~\ref{tab:SI_nsp_details} in SI Text 4 we list the $\alpha_0$ and $\alpha$ for all and time periods.
It's instructive to compare the values for supplier turnover in Fig.~\ref{fig:spontaneous_link_generation}b with the rates for customer turnover in Fig.~\ref{fig:link_decay}c. Some sectors have high link-decay (as supplier) and high $\langle N^{s+}\rangle $  (as buyer), these sectors, for example ``I ~ Tourism" have high turnover up- and downstream. Sectors with low link-decay (as supplier) and high $N^{s+}$ (as buyer), for expample ``E - Water \& waste collection/treatment" have stable customers, but switch suppliers often.

\subsection*{Link entry -- suppliers}
Next, we characterize the new links from the supplier perspective. We do this in 2 steps. First, we analyze to what extent some sectors are more likely to link to each other by calculating the conditional probability for a customer in sector $s_2$ to link to a supplier in sector $s_1$, summarized in the \textit{supplier attachment probability} matrix $\Pi(s_1,s_2)=p(s_1 | s_2)$, shown in Fig.~\ref{fig:link_attachment} a. High (low) linking probabilities are shown as light (dark) colors. The matrix is relatively sparse, meaning that most new links are focused on a few sector combinations only. The pronounced diagonal shows that firms are likely to link to firms in their own sectors. A few vertical lines are prominently visible, e.g.,  "G46 - Wholesale", ``D35 - Electricity, gas, steam and air conditioning supply", and ``L68 - Real estate activities". 
These sectors are important suppliers to many other sectors. There are also two faint blocks visible, corresponding to manufacturing (1-43 or A-F, upper left) and service (45-99 or G-Z, lower right) sectors. The average within-block similarity is  $\langle \Pi_{[A-F],[A-F]}\rangle = 0.013$ and $\langle \Pi_{[G-Z],[G-Z]}\rangle = 0.019$, the average off-block similarity $\langle \Pi_{[A-F],[G-Z]}\rangle = 0.011$ and $\langle \Pi_{[G-Z],[A-F]}\rangle = 0.004$. We test that the difference is statistically significant using a Mann-Whitney U test, rejecting the null hypothesis that the blocks are sampled from the same distribution ($p<0.001$). We use the non-parametric test because neither the diagonal nor the off-diagonal similarity values follow a Gaussian distribution.

Second, we ask how likely will a new link connect to a supplier with degree $k$ in the selected sector, $s_2^*$? We answer this by estimating the function that determines the probability of attracting new out-links in a given sector. This function is only based on the current degree of the node, and we will call it the \textit{attachment kernel}, $A^k(k)$. We measure it by adapting the method used in~\cite{newman2001clustering} and applying it to the monthly data in period A. The probability that a new link attaches to a node with degree $k$ is now given by $P(k) = \frac{1}{\tilde{\mathcal{N}}}A^k(k)n(k,s_2^*)$, where $n(k,s)$ denotes the number of nodes with degree $k$ in sector $s$, and $\tilde{\mathcal{N}}$ the normalization. The attachment kernel $A^k(k)$ can be estimated from a histogram where we count the number of new suppliers per degree bucket weighted by $\frac{1}{n(k,s)}$. We use linear buckets of size $\Delta k = 1$.
In Fig.~\ref{fig:link_attachment} b we plot the number of new suppliers per degree bucket per month for period A, weighted by $\frac{1}{n(k,s)}$. We fit an attachment kernel of type $A^k \propto k^\beta$ using OLS on the log-variables and find slightly super-linear attachment, $\beta = 1.079 \pm 0.018$, we report the 90\% CI.
In Fig.~\ref{fig:link_attachment} c we fit $A^k$ for different NACE industries, unveiling a slight bias towards super-linear attachment with a few exceptions in C, D, J, K, O, and U, see also Tab.~\ref{tab:SI_Ak_details} for sector names and the results for the other time periods.

\begin{figure*}[t]
\centering
\includegraphics[width=0.90\textwidth]{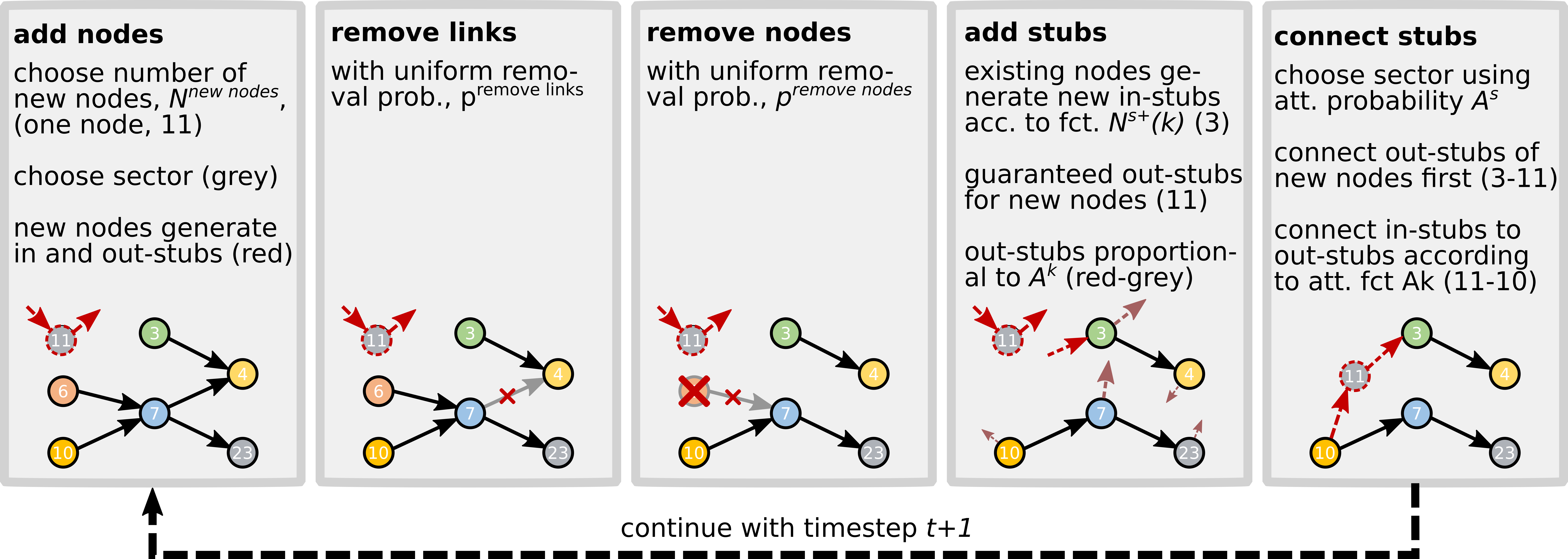}
  \caption{schematic view of the SCN generating model. It evolves from one timestep to the next in the following five steps:
  1. add new firms. The number of firms is sampled from a Poisson process and the sector is drawn from the empirical sector distribution. New nodes are added to the network with a number of in- and out-links, $k^{in}$ and $k^{out}$ sampled from the empirical joint distribution, $p(k^{in}, k^{out})$.
  2. remove links with uniform probability, $p^{\mathrm{remove links}}$.
  3. remove firms with uniform probability, $p^{\mathrm{remove nodes}}$. All their links are eliminated. We remove all isolated nodes that lost their links in this link elimination.
  4. add stubs -- spontaneous link creation. Existing nodes generate new in-stubs according to $N^{s+}(k)$.
  5. connect firms. In-stubs are connected by first choosing a sector to attach them to using the sector attachment probability, $\Pi_{s_1 s_2}$, and then picking a node with probability that is proportional to the attachment kernel, $A^k(k)$. However, before connecting to nodes based on $A^k$, links are connected to out-stubs of new firms. This guarantees that firms keep attached to the SCN.
  }
  \label{fig:model_steps}
\end{figure*}

\subsection*{Generative model for the evolution of SCNs}
\begin{figure*}[ht]
    \centering
    \includegraphics[width=0.99\textwidth]{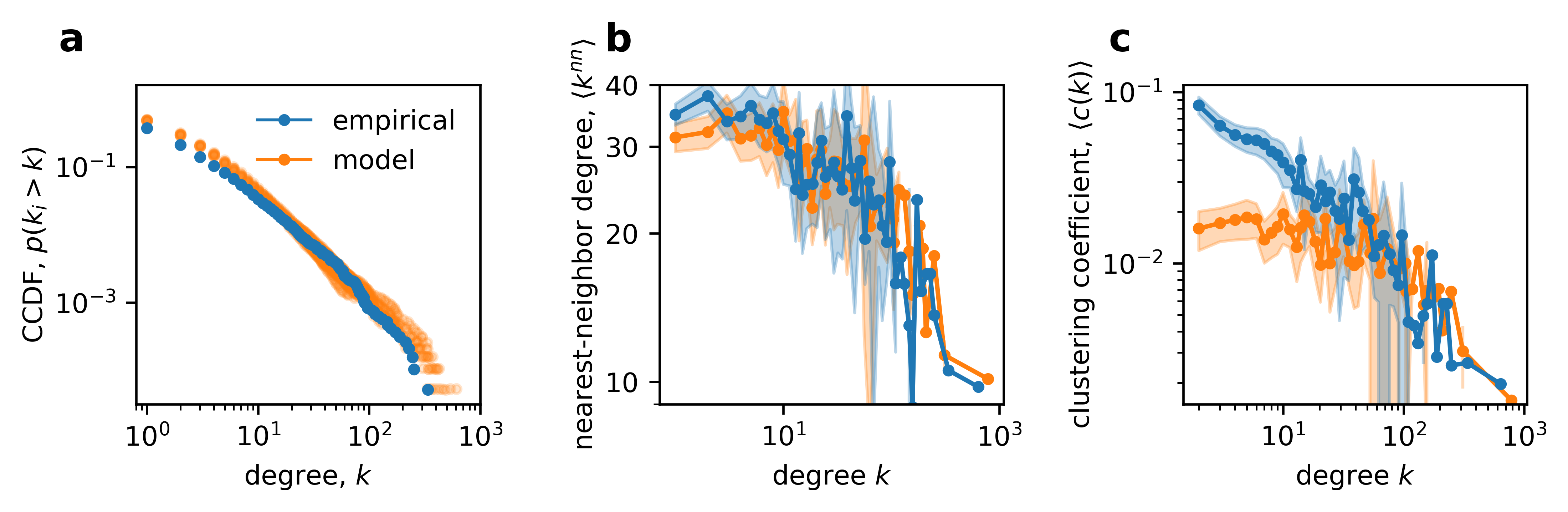}
    \caption{Model results of structural properties of SCNs.
    (a) Counter cumulative degree distribution, $p(k_i \geq k)$ for snapshots generated by the supply network generating model (orange) 
    and the empirical degree distribution in January 2017 (blue). The distributions are very similar.
    (b) Average nearest neighbor degree, $\langle k^{nn} \rangle$, for the empirical (blue) and model (orange) networks using linear bins for  
    $k\le 10$ and log bins for $k > 10$, the shaded area denotes the  90\% CI. The dis-assortativity structure of both networks is 
    practically identical.
    (c) Average local clustering coefficient $\langle c_i \rangle_{i | k_i \in [k,k+{\rm bin}]}$ as a function of degree for the empirical (blue) and 
    model (orange) networks (using same binning and error). The hierarchical structure reflected by the power law decay of the clustering 
    coefficient as a function of the degree is again well captured especially for large degrees, but is somewhat underestimated for low degrees.
    }
    \label{fig:model_results}
\end{figure*}

We are now in the position to use the previous estimates to calibrate a simple statistical generative model to the actual Hungarian SCN. The model is initialized with a set of firms (nodes), $\mathcal{N}$, and a set of links, $\mathcal{L}$. Here, we initialize the model with one snapshot of the empirical network, i.e., Jan. 2017. The network follows the topology shown in blue in Fig.~\ref{fig:model_results} and contains 18,805 nodes. Note, this network is significantly smaller than shown in Fig.~\ref{fig:concept_figure}d, because here we use only firms and links in the `persistent network', see Materials and Methods. We initialize the model with the parameters shown in the first 5 rows of Tab.~\ref{tab:model_results}. We perform the following five steps at every timestep; for a schematically overview, see Fig.~\ref{fig:model_steps}.

\begin{enumerate}
\item {\em Add new firms.} The number of new nodes is drawn form a Poissonian distribution with mean $N^{\text{new nodes}}$, as  shown in Fig.~\ref{fig:entry_exit_rates}a. Every node enters with an in-degree, $k^{in,0}$, and out-degree, $k^{out,0}$, that are sampled from the empirical joint distribution of in- and out-degree upon entering, $p(k^{in,0}, k^{out,0})$, for details see SI Text 6. For  every newly created firm, $i$, we draw its sector, $s_i$, from the empirical sector-distribution, $p(s)$  as in \eqref{eq:entry_sector_probability}.

\item {\em Spontaneous link removal.} With probability, $p^{\text{remove link}}$, (see \eqref{eq:link_removal_prob}) every existing link in the SCN is eliminated, regardless of any features of the nodes or link weight. 

\item {\em Firm removal.} Firms are removed with a uniform removal probability, $p^{\text{ex}}$  (see \eqref{eq:node_removal_prob}. All links connected to these firms are eliminated. We then also remove all those nodes that have become isolated after this link-removal step.

\item {\em Spontaneous link creation.} In- and out-stubs are added to every existing firms, $i$, that determine the number of new in- and out-links that the firm will acquire in this timestep. In-stubs represent to how many new suppliers each firm will connect to; out-stubs are proportional to the probability that an in-stub connects to a given supplier, $j$, i.e. it's \emph{attachment kernel} $A^k(k)$. The number of in-stubs is drawn form a binomial distribution with mean $\langle N^{s+}\rangle(k_i)$, (see \eqref{eq:new_links_per_cust} in section ``Link entry - customers") as shown in Fig.~\ref{fig:spontaneous_link_generation}. The out-stubs of every firm, $j$, are proportional to its \emph{attachment kernel}, $A^k$, as calculated in Sec. ``Link entry - suppliers" and shown in Fig.~\ref{fig:link_attachment}b.
Both quantities are not sector specific.

\item {Connecting the firms.} We connect every in-stub to a supplier in a two step process: 
(i) We choose the sector of the target firm, $j$, (supplier) with probability $\Pi_{s_i,s_j}$, and 
(ii) then connect the in-stub to a firm in $s_j$ with probability, $p(j) = A_j / \sum_{k \in s_j} A_k$; see Sec. ``Link entry - suppliers".
The process ends when all in-stubs are connected to a supplier. To make sure that every new firm reaches its determined in- and out-degree, we connect to them first, until they have acquired $k^{in,0}$ and  $k^{out,0}$ customers. 

\end{enumerate}

The only size dependent mechanisms are the ones determining in- and out-degree in step 4 and 5. The only step depending on the firm's sectors is when the stubs are connected in step 5. New nodes are added with pre-determined size and sector, and both firm and node exit isn't influenced by neither, size, nor sector.

\subsection*{Model results} 
While the number of firms in the real network is growing by 3.3 \% every year, we prefer to simulate a stationary economy and slightly increase the node removal probability $p(\text{node exit})$ such that we remove the same number of nodes that is added on average.
In SI Fig.~\ref{fig:SI_model_timeseries_NL_2017} we show the timeseries of number of firms and links.

We initialize the model with the empirical network topology of January 2017 and run it for 500 time steps. A timestep is calibrated to correspond to one month. After this initial phase we take a snapshot of the model SCN at every 200 timesteps for further analyses. In total we analyze 10 such snapshots. The model captures the network size dynamics with respect to the number of firms, $N$, and links, $L$; see Tab.~\ref{tab:model_results}. In SI Fig.~\ref{fig:SI-model_DeltaNL} we show the monthly change of firms and links, $\Delta N$ and $\Delta L$, respectively. Both quantities show a negative skew with a long negative tail; since we remove isolated nodes, the spontaneous exit of high-degree nodes can cause cascades of exits starting from its neighbors. 

The firm- and link-volatility in the model as measured by the 12-month sample variance, $\sigma^2_{N}= 36,067$ and $\sigma^2_{L}= 1,558,618$, respectively, underestimates the empirical variance in both periods, which are $\bar \sigma^2_{N}= 704,671$ and $\bar \sigma^2_{L}= 24,121,146$, respectively. This is a consequence of the fact that model doesn't take empirical seasonality into account.

We continue with describing structural network features that emerge from the model. In Fig.~\ref{fig:model_results} we show the degree distributions, $p(k_i > k)$  (CCDF), of the snapshots (orange) and compare them to the empirical degree distribution of January 2017 (blue). Note the agreement across almost 3 orders of magnitude. We find average degrees of $\langle k^{model}\rangle = 3.06$ and $\langle k^{emp}\rangle = 2.81$. The model also captures the differences the empirical in- and out-degree distributions; see SI Text 7.

SCNs are known to be dis-assortative, meaning that high-degree firms tend to be linked to low-low-degree firms and vice-versa~\cite{fujiwara2010large,bacilieri2023firm}. We confirm this assortativity structure in Fig.~\ref{fig:model_results} b by plotting the average nearest neighbor degree, $\langle k^{nn} \rangle = (1/k_i) \sum_{j \in \mathcal{N}_i} k_j$, as a function of degree $k$, where $\mathcal{N}_i$ is the set of direct neighbors  of $i$. The empirical network is the blue line, the ten model snapshots are the orange lines. The model reproduces the pronounced dis-assortative structure very well for all degrees.

A decaying clustering coefficient $c_i$, when plotted as a function of the degree $k$, indicates the presence of  nontrivial hierarchy in complex networks \cite{ravasz2003hierarchical}. It is defined as $c_i = 2t_i/k_i(k_i -1))$, with $t_i$ the number of triangles $i$ is involved in. In Fig.~\ref{fig:model_results} c we plot $\langle c_i \rangle_{i | k_i \in [k,k+{\rm bin}]}$ as a function of degree using bins described in the caption. The model appears to accurately reproduce the hirarchical structure of SCNs especially for large $k$, but deviates for low degrees. 
This might be because the model lacks a mechanism for triadic closure as proposed by other authors~\cite{chaney2014network,carvalho2014input}, which seems to be most important for low degree firms.

\begin{figure}[t]
    \centering
    \includegraphics[width=\columnwidth]{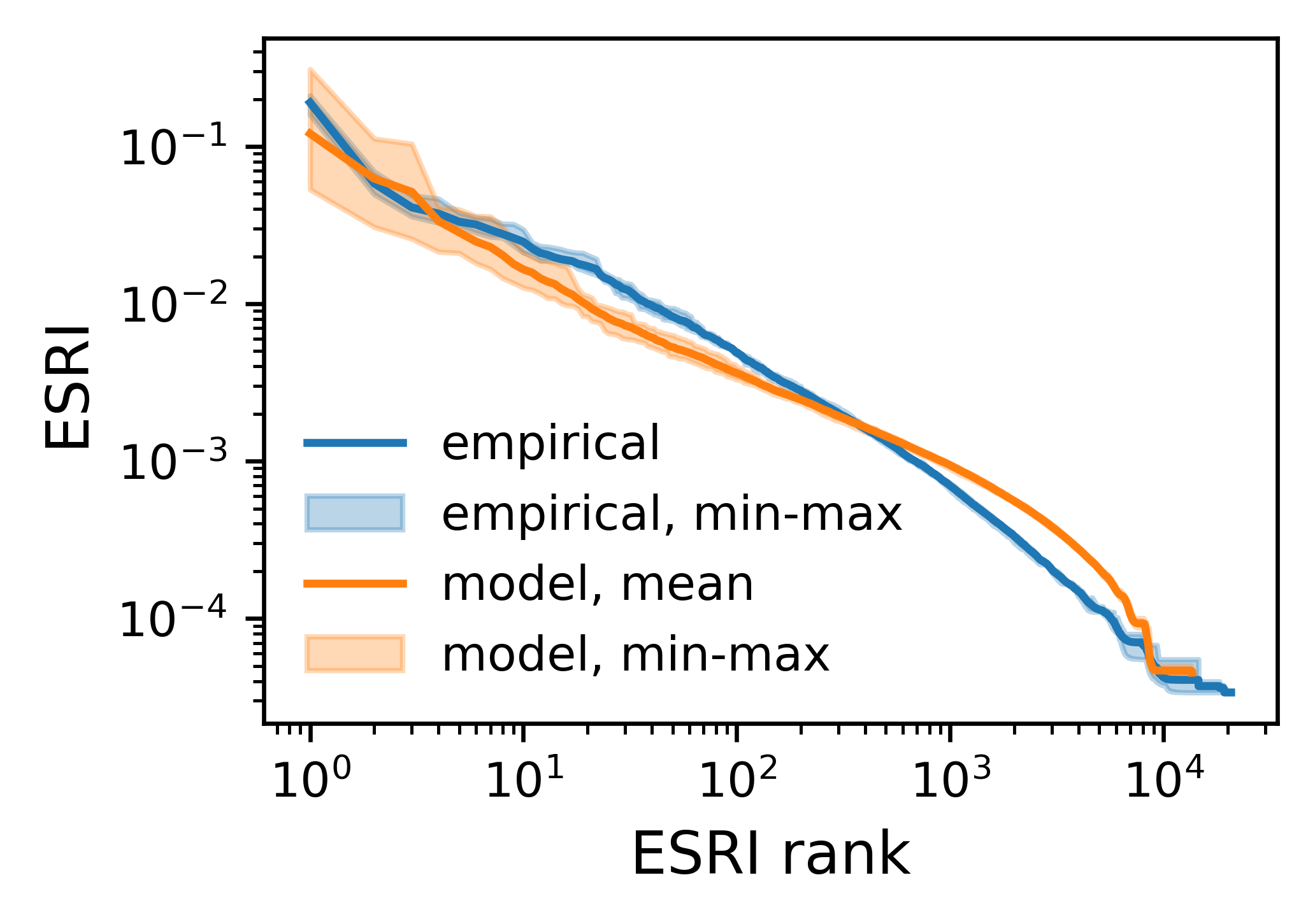}
    \caption{Economic Systemic Risk profile as measured with ESRI, from empirical data (blue) and ten model networks from ten snapshots 
    (orange). The ESRI value of every node is plotted against its ESRI rank. The empirical and model networks have remarkably similar 
    systemic risk profiles across 4 orders of magnitude.
    }
    \label{fig:model_results_ESRI}
\end{figure}

\begin{table*}[t]
	\centering
	\caption{Model calibration and results. The first part of the table shows the calibration values used for each time period (see Fig.~\ref{fig:concept_figure}). We show the average number of firm entries, $\langle N^{entry} \rangle$, the node exit probability, $p(\mathrm{node exit})$, the intercept and exponent of the link generation scaling function $\langle N^{s+}\rangle$, $\alpha_0$ and $\alpha$, respectively, the Attachment kernel exponent $\beta$, and the link removal probability $p^{term}$. In the second part of the table we compare empirical and modelled quantities, in particular, we show the number of nodes, $N$, number of links, $L$, the average degree, $\langle k \rangle$, and the respective variances, $\hat{\sigma}^2(N)$ and $\hat{\sigma}^2(L)$.}
	\label{tab:model_results}
	\begin{tabular}{l|rr|rr|rr}
		
		& \multicolumn{2}{c}{Period A} & \multicolumn{2}{c}{Period B} & \multicolumn{2}{c}{Period C} \\
		& empirical & model            & empirical & model            & empirical & model \\
		\midrule
		\multicolumn{7}{c}{Calibration quantities} \\
		\midrule
		$\langle N^{entry} \rangle$ &    357 & = &    605 & = &    904 & = \\
		$p(\mathrm{node exit})$     & 0.0049 & = & 0.0051 & = & 0.0046 & = \\
		$\alpha_0$                  & 0.0108 & = & 0.0118 & = & 0.0160 & = \\
		$\alpha$                    & 1.0369 & = & 0.9955 & = & 0.9711 & = \\
		$\beta$                     &   1.08 & = &   1.08 & = &   1.06 & = \\
		$p^{term}$                  & 0.0214 & = & 0.0195 & = & 0.0338 & = \\
		
		\midrule
		\multicolumn{7}{c}{Modeled quantities} \\
		\midrule
		$N$                 & 20059.9 & 18339.9 & 33971.6 & 33700.6 & 41817.2 & 41646.6 \\
		$L$                 & 28177.3 & 28132.2 & 59349.9 & 54517.0 & 79474.1 & 64143.9 \\
		$\langle k \rangle$ &    2.81 &    3.06 &    3.49 &    3.24 &    3.80 &    3.08 \\
		
		$\hat{\sigma}^2(N)$  &   704671 &   36067 &   682905 &   198381 &  2825670 & 398510 \\   
		$\hat{\sigma}^2(L)$  &  2412146 & 1558618 &  3858995 &   894264 &  19535266 & 1057220 \\     
		\bottomrule
	\end{tabular}
\end{table*}

{\bf Systemic risk profiles.}
After having examined local network structures of the modeled SCN, we now turn to the question if the model also captures realistic structures that are relevant for spreading of defaults. For this we turn to the Economic Systemic Risk Index (ESRI) as developed in~\cite{diem2022quantifying}. These structures are know to operate at a semi-local or meso-level. The ESRI value of a firm quantifies the economic damage that its failure would immediately cause to the entire SCN, taking into account the network structure and production functions; for details see Materials \& Methods. ESRI is highly sensitive to the structure of the SCN, in particular correlations between network- and sector-structure~\cite{diem2024estimating}. We calculate the ESRI for every firm in the empirical network of January 2017 and plot the rank-ordered distribution, from the highest to the lowest, in Fig.~\ref{fig:model_results_ESRI} (blue line). Note, here that the empirical network of stable connections exhibits no plateau of high systemic risk firms that was observed in annual data~\cite{diem2022quantifying}. We repeat the procedure for the ten model network snapshots and plot the mean ESRI for every rank (orange line) as well as the respective minimum and maximum ESRI values (orange shaded area). The ESRI of the 20 empirically most risky firms lies well within the min-max bounds of the model runs. For ranks higher than 20 the model underestimates the emprical ESRI, before crossing at around rank 400, where the empirical ESRI becomes smaller than the model estimate for the least risky firms. The network generative model manages to reproduce the ESRI well for the most risky nodes, however,  the scaling exponent in the rank ordered plot is somewhat lower than in the empirical network (from rank $10-1000$).

\textbf{Results for different time periods.}
The model calibration was derived for period A. We perform the same calibration also for periods B and C and compare the results in Tab.~\ref{tab:model_results} and SI Text 7. 

We list the calibration quantities in Tab.~\ref{tab:model_results}.  The parameters determining firm and link entry/exit rates ($\langle N^{entry} \rangle$, $p(\mathrm{node exit})$, $\alpha_0$, $p^{term}$) are all higher in period B and C, respectively, because the network is larger and features more small firms and links that are more volatile.
We characterize the network with several key quantities. The model in period B and C produces slightly sparser networks than empirically observed, as shown by 5\% and 20\% lower values for $L$ and $\langle k \rangle$, respectively. The underestimation of the number of links and, consequently, the average degree, increases with the lowering of the reporting threshold, highlighting that when more volatile low-value links are included the sparse model calibration doesn't reproduce the full network structure. We also provide the variance of the $N$ and $L$ timeseries, $\hat{\sigma}^2(N)$ and $\hat{\sigma}^2(L)$, respectively. In all periods. For both timeseries we underestimate the variances because we don't model the empirical seasonality.

\section*{Discussion}

We present a detailed statistical analysis of the four processes that govern the dynamical evolution of supply chain networks based on monthly VAT data from Hungary. We presented a minimum statistical model, in the sense that we try to use rates and marginal distributions whenever possible, and only employ joint distributions when unavoidable. This results in a network generative model that is calibrated with the data and  reproduces local structural network characteristics of the empirical SCN, including systemic risk profiles (that are based on non-local information).

We find rates of 2.1\% node turnover (exit) and 4.6\% link turnover (exit) per month. This means 25\% (exit, 28\% entry) node turnover and 55\% (exit, 61\% entry) per year. This is a somewhat unexpected result. While it is known that firms exit and enter at rates of 8 \% to 13 \% \cite{eurostat_sbs}, the amount of turnover in links is less known and is striking. All of the economy is  tremendously dynamical, it rewires completely every few years. This is an indication why it is resilient, it is able to reconfigure in responses to shocks because it is used to constant rewiring. Previous work has emphasized the relevance of the restructuring process of the SCN for the economic success and decline of firms~\cite{coe2005internationalization}, the resilience of production networks~\cite{mari2015adaptivity}, and mitigating economic shock spreading~\cite{diem2022quantifying,konig2022aggregate}. The adaptive nature of SCNs has been emphasized~\cite{choi2001supply}. Our work contributes to this theoretical literature by providing an empirical description of the network restructuring processes.

We fit the node- and link entry- and exit rates with seasonally adjusted Poisson processes. This is found to describe the entry rates  of firms and links well without considering sector or link-weight information. Of course, model results could be improved by incorporating the empirical heterogeneities in entry- and exit rates rates by sector and firm size, see SI Text 1. We focused on a minimal model, and see this as an avenue for future work.

We investigate the link exit process in more detail and find that the process is reasonably described by a constant decay rate, suggesting a process that is to first order memoryless, see SI Text 3. The decay rate depends on the link weight and varies strongly for links in the lowest weight quintile. This can introduce a bias in the network structure, because large link weights (sales) are typically associated with large firms --- large sales in the network are orders of magnitude larger than those of small firms. The decay rate also varies strongly across the industrial sectors of the suppliers, reflecting the fact that time horizons vary for different economic activities.

We study link entry by characterizing both the customers and suppliers. The number of new suppliers per firm and per month can be described by a power law of the degree,  $\langle N^{s+} \rangle = 0.011 k^{1.04}$, that is compatible with a linear dependence. We find that the linear pre-factor can be different for different sectors. This is intuitive, since as links get removed with a constant rate, larger firms need to generate new links proportional to their degree to maintain their size. After determining the number of new in-links a customer generates per month (changes of suppliers), we study the characteristics of the suppliers they attach to. We find that the supplier attachment matrix, containing the conditional probability for a customer in sector $s_2$ to attach to a supplier in sector $s_1$, $\Pi_{s_1,s_2}$, is very sparse with the exception of the diagonal, i.e. the firm's own sector, and a few `star' sectors, to which most other sectors frequently connect. These 'star' sectors that supply to most other sectors include the NACE categories ”G46 - Wholesale”, “D35 - Electricity, gas, steam and air conditioning supply”, and “L68 - Real estate activities”.

In a given sector, firms tend to connect to large suppliers, a process well known as preferential attachment~\cite{barabasi1999emergence} (PA).  We find super-linear preferential attachment with an average attachment kernel scaling exponent $\beta = 1.08$. However, $\beta$ can vary strongly between sectors. For example sectors 
`D - Electricity, Gas, Steam and Air Conditioning S...', 
`K - Financial and Insurance Activities', and
`O - Public Administration and Defence' feature pronounced sub-linear PA, 
while 
`F - Construction', 
`I - Accommodation and Food Service Activities', and 
`L - Real Estate Activities' feature strongly super-linear PA.
We find, however, clear evidence of preferential attachment across all industries with $\beta > 0$ for every NACE section.
Not including this sectoral heterogeneity will lead to wrong estimates of the tail exponents of industries.

We formulate and calibrate a simple network generative model that mimics the dynamics of a national SCN. As inputs it takes the rates and probabilities and probability distributions that were described so far. We test the model by comparing the structural network characteristics that emerge. We take snapshots of the emerging SCNs once in a stationary state and compare the derived characteristics directly with the empirical counterparts. The model is kept as simple as possible, ignoring the link-weight, sector, {and seasonal} dependency of the input parameters, whenever possible. Certainly, it possible to estimate link-weight, sectoral, and time dependent rates and probabilities in greater detail given enough data. Using these would then allow the model to capture the sectoral and seasonal network structure. This is subject to future work.

The generative model reproduces the number of firms as well as the in- and out-degree distributions.
We find tail exponents of $2.653\pm0.087$ and $2.620\pm0.76$ for the modeled in- and outdegree distributions, respectively. These values are slightly higher, but of similar magnitude as the exponents of the empirical in- and outdegree distributions, $2.383 ± 0.081$ and $2.452 ± 0.084$, respectively (we report 90\% CI intervals). Note, that we report significanlty larger tail exponents than what is commonly found in the literature~\cite{bacilieri2023firm}. This is due to the fact that we consider  `stable' connections only, see Materials and Methods. Previous models have not modelled the in- and out-degree distribution at the same time~\cite{atalay2011network}.

The model further reproduces the assortative mixing of the emprical SCN realistically,  meaning that the network is dis-assortative, i.e. small firms tend to link to large firms and vice versa. The dis-assortative network structure is highly consistent across the literature~\cite{bacilieri2023firm} and distinguishes SCNs from other types of networks as for example social networks~\cite{reisch2022monitoring}.

We find some deviations in the clustering behavior; for small degrees below twenty the empirical average local clustering coefficient is about a factor 2 larger than those from the model. Even though clustering coefficient is computed for undirected triangles, we checked whether the problem originates from the triangles where firms connect to suppliers of suppliers or firms connect to customers of their customers, but find no clear difference. This might arise from the fact that the model does not incorporate an explicit mechanism for triadic closure.
In the literature the tendency to connect to suppliers of suppliers, which leads to closed triangles and higher clustering, has been proposed as a  mechanism in SCN formation \cite{chaney2014network,carvalho2014input}. 
Our model misses this, because the attachment mechanism is based only on first-order properties of the node itself, not higher-order properties, such as the characteristics of a nodes' neighbors. One could incorporate such mechanisms straight forwardly by making the attachment kernel depend not only on a node itself, but also on its neighbors.

Another potential explanation for the underestimation of the local cohesiveness is that our model does not consider the geographical proximitiy of firms, which is known to play an important role for supply link formation~\cite{bacilieri2023firm}. If firms tend to connect to firms within their geographical region, reciprocal links and triads are more likely. Firms with many suppliers will have to source from outside their geographical region and the difference in clustering vanishes.

The model also is also able to reproduce the essence of the empirical \textit{economic systemic risk profile}. There a few firms with a high ESRI  value are observed (sytemic risk core; see \cite{diem2022quantifying}) and a power-law decay in the rank ordered ESRI distribution covering a similar range of values as reported in the literature (see SI Text 8 and \cite{diem2022quantifying}). This is somewhat unexpected, since ESRI is highly sensitive to correlations between network- and sector-structure~\cite{diem2024estimating}. 

Empirical and modelled ESRI values are of similar magnitude, their tails drop with a different slope in the rank ordered distribution. Note,  that the empirical ESRI profile doesn't show a plateau of high systemic risk firms that was observed in yearly aggregated data~\cite{diem2022quantifying, reisch2022monitoring}. The filtering of links with at least three connections within six months causes the high systemic risk core to fracture and no plateau emerges; for details, see SI Text 8.

To get a feeling for the relevance of overfitting, we also calibrate a generative model with the data from the other time periods. We find it robust with respect to network size and other parameter combinations, however, the model underestimates the number of links and, consequently, the network density. This is because for the later periods the number of new suppliers scales sub-linearly. With uniform link removal probability this means that large nodes loose more links than they can replace. Hence, for the later periods it would be more relevant to model the size dependent link removal rate, as shown in Fig.~\ref{fig:link_decay}b. The sample variance of firms and supply links is underestimated because the model does not explicitly model seasonality. 

Two network generative models for production networks were developed previously. Reference \cite{atalay2011network} deals with the question whether the buyer–supplier network of the US economy is purely scale-free. They derive model parameters mostly from micro-data, but fit one last parameter to the in-degree distribution using a maximum likelihood estimator. In comparison, in our present model all involved processes and  all parameters are derived and quantified on the micro-level, there are no free parameters. Further, our model simultaneously reproduces the in- and out-degree distribution of the Hungarian economy, compared to only the in-distribution in \cite{atalay2011network}.

In~\cite{ozaki2024integration} a two-layer model is proposed for the Japanese production network, combining a generative network model for the network topology with a diffusion model for the link weights. Only firms that enter the network form new links using preferential attachment; established firms grow by merging with other nodes. Model parameters are fitted to the emergent network properties such as the degree distribution, firm degree-growth rates, and the sales growth rate. In the presented model we take the opposite direction, by directly investigating firm behavior on the microscopic level and using a generative model to understand emergent properties which can be immediately compared with the empirical data. With our model we cannot confirm the results  of~\cite{ozaki2024integration} because, for example, their model describes the data best if 37\% of firms are subject to mergers, and --due to a model assumption-- the same fraction of is also subject to splitting. In Hungary these rates are much lower with values around 4 \% in the years between 2015 and 2021.

Our study is subject to four obvious limitations. First is timing issues in the data quality. We do not know if the transactions occurred at the same time as the recorded VAT payments. The VAT reporting guidelines require firms to report transactions in the months they occurred, however, misreporting and trade credit, which can cause delays in payment of up to 120 days~\cite{proselkov2024financial}, can cause distortions in the sequence of link formation. We expect this effect to be small because if these reporting delays remain approximately constant, our parameters should not be affected. To accurately model link weights we would need to consider production capacities of firms and consistency between in- and output quantities, according to firm's production functions, which is beyond the scope of our current technical capabilities. Third, we only consider the network of {\em stable} supply links to ensure a clear meaning of the beginning and ending of a buyer-supplier relationship (with having physical production networks in mind). This means that we discard the role of volatile and short-term links that might be essential, such as investment goods that don't occur as regular payments, but as one-off purchases. Fourth, several micro-economic mechanisms for link formation in SCN and production networks have been proposed \cite{oberfield2018theory,gualdi2016emergence,chaney2014network, carvalho2014input,mundt2021formation,balland2012proximity,konig2022aggregate}. Here we study the `effective' outcome of these mechanisms only. 
For example, the economic processes underlying the preferential attachment include (among others) the fact that only some suppliers produce the products needed~\cite{konig2022aggregate}, the social network underlying the formation of business ties~\cite{chaney2014network,carvalho2014input}, and of course price differences~\cite{gualdi2016emergence}. The situation might become more transparent when  more data on the involved firms could be included, such as geography, productivity, sales, etc.~\cite{mungo2023reconstructing}, which is impossible for us at this stage. In future work higher order processes could be added to the model, such as triadic closure mechanisms, firm splitting or merging, or introducing explicit seasonality.

Data of national supply chain networks on the firm level remains to be a hot topic\cite{pichler2023building}. Typically, it is highly protected and can not shared freely. We see the value of this work not only in a comprehensive descriptive statistics showing a massive turnover in the economy, both in terms of firms and their supply links, but also to make a `digital twin'  available to a wider community in form of the presented model; for the source code, see \url{https://github.com/treisch/scn_generative}.

\section*{Materials and Methods}
\label{sec:matmeth}

\subsection*{Data} 
We use VAT data provided by the Hungarian central bank. Since 2014 Hungarian firms need to report their suppliers with whom they exceeded a threshold of 1 million Hungarian Forint (HUF) tax content in the reporting period (period A). In July 2018 the threshold was lowered to 100,000 HUF tax content and reporting is evaluated at the invoice level (period B). In June 2020 the reporting threshold was set to zero and firms needed to report all partners and invoices, irrespective of transaction volume (period C). In 2014, the first year of granular VAT reporting in Hungary, data quality is poor and we do not consider this year in the analysis. Hungarian VAT rates range from a 27\% base rate to a 18\% and 5\% reduced tax rate for certain foods, pharmaceuticals, etc., and there is a 0\% rate for public transport~\cite{vatratesHUN}. Firms that had a net VAT remittance payments lower than 250,000 HUF in the year before the previous and less than 50 million HUF revenue without taxes, can report their suppliers on an annual basis. Firms exceeding these thresholds but that are below a net VAT remittance of 1 million HUF report their suppliers on a quarterly basis, the remaining firms report their transactions monthly. Firms that exceed one of these thresholds within a given year have to change to the reporting frequency that applies. The data contains information on firms transactions aggregated to monthly, quarterly, or annual levels, based on the reporting firm's size, as well as information on the firm's industry and size. We focus on monthly reported transactions which comprise on average 85.6\% of the reported production volume.

\subsection*{Stable supply links}
For our analysis we are interested in sufficiently stable supply connections and not in one-time purchases, even though these might be relevant. We define the existence of a supply link between two firms if there occur at least three transactions within a time window of six consecutive months, see SI Fig.~\ref{fig:SI_filtering_procedure}. The time of a link-entry is the month of the first transaction that occurs in a time period where this condition is fulfilled, for the subsequent timesteps (months) we mark the link as `existing', the time of link-exit is the month following the month in which the last transaction took place (where the condition is satisfied). The procedure removes about 77.9\% of all links in period A, which accounts for about 18.5\% of the overall purchase volume. The time a firm enters the SCN is recorded as its first overall occurrence; the time a firm exits by its last overall occurrence. 

\subsection*{ESRI}
We employ the systemic risk measure, ESRI, as described in~\cite{diem2022quantifying}. For every firm, $i$, it measures the immediate relative reduction of production in the SCN as a consequence of the  firm $i$'s default. The algorithm requires the SCN information, an estimate of the production functions, and one has to specify the relative `essentialness' of sectors, for which we use a survey among industry experts, described in~\cite{pichler2022forecasting} and~\cite{diem2024estimating}. We use two-digit NACE codes for the industries. Note that here we do not consider link weights or the cost/revenue correction as in previous works. In SI Text 8 we analyze the effects of these parameter choices. The source code for ESRI is available \url{https://github.com/ch-diem/misestimation_from_aggregation}.

\bibliography{pnas-sample}

\section*{Acknowledgements}
We thank Christian Diem, Janos Kertesz, and Francois Lafond for helpful discussions. T.R. thanks the Hungarian National Bank for hosting his research visits. The project was supported by Austrian Science Fund FWF I5985 and P33751, 
Hoschuljubiläumsstiftung der OeNB 18696,
and Austrian Research Promotion Agency FFG 873927.

\section*{Author contributions statement}
ST and TR conceptualized the work. TR and AB curated and prepared the data. TR and ST analyzed the data and wrote the paper.

\section*{Additional information}
\textbf{Competing interests} The authors declare no competing interests.


\newpage
\onecolumngrid  
\FloatBarrier
\section*{Supplementary Information}

\renewcommand{\figurename}{SI Fig.}
\setcounter{table}{0}
\renewcommand{\thetable}{S\arabic{table}}%
\setcounter{figure}{0}
\renewcommand{\thefigure}{S\arabic{figure}}%

\subsection*{SI Text 1: Disaggregated turnover}

Industries organize their supply chains in different ways and turnover rates for links are not the same for all sectors and firm sizes. Here, we discuss entry and exit rates by supplier and customer sector and size.

\begin{figure*}[htb]
	\centering
	\includegraphics[width=0.8\textwidth]{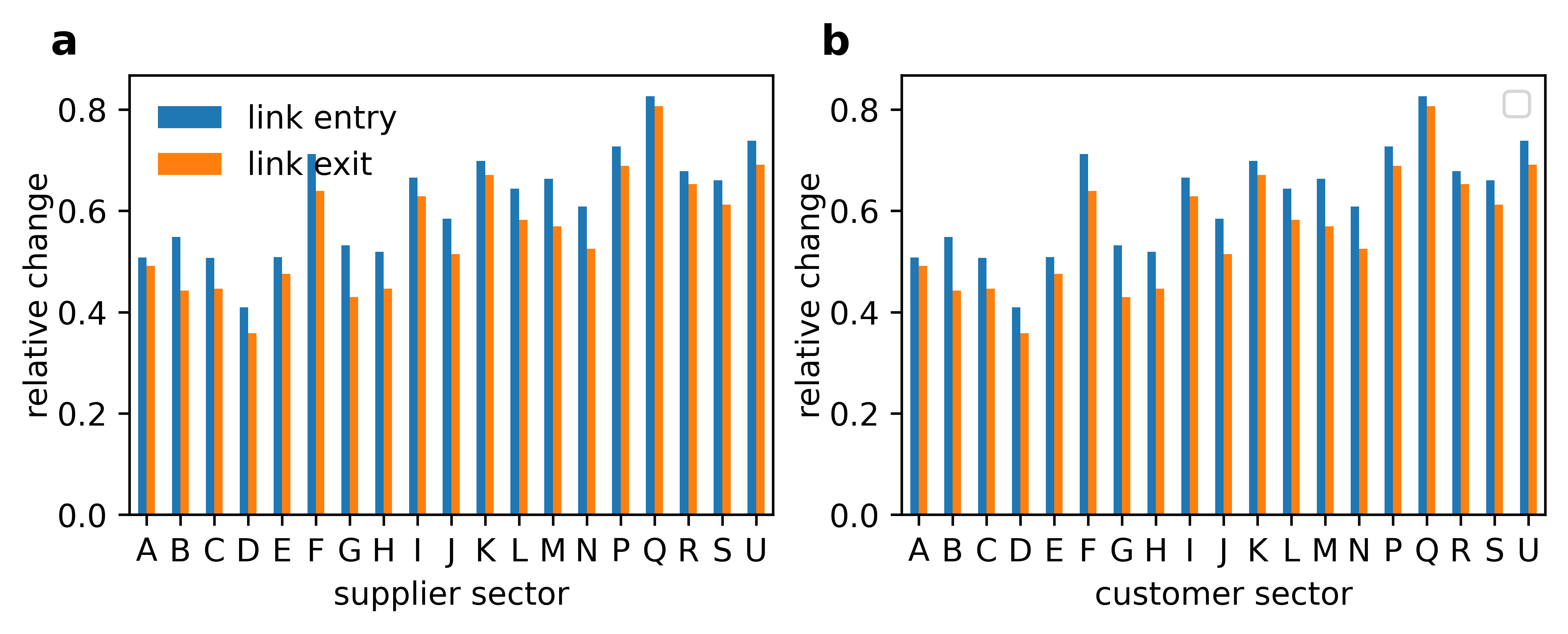}
	\caption{Link turnover in period A by supplier and customer industry.
		Relative link entry rates are shown in blue and exit rates in orange by (a) 
		supplier sector and (b) customer sector. Turnover rates vary strongly between industries and are lower for manufacturing sectors except construction (A-E) than for service sectors (I-U).}
	\label{fig:SI_rel_entry_exit_sector_1d}
\end{figure*}

In Fig.~\ref{fig:SI_rel_entry_exit_sector_1d}a we show relative link entry (blue) and exit (orange) rates by supplier sector for period A. 
The sectors with the highest entry rates are NACE sections Q, P, F, K (see Tab.~\ref{tab:SI_NACE_letters} for sector descriptions); the sectors with the highest exit rates are sectors Q, P and K. 
For most supplier sectors entry rates are higher than exit rates, because the entire network is growing. Net growth is highest for B, G and M.
In Fig.~\ref{fig:SI_rel_entry_exit_sector_1d}b we plot the relative link entry (blue) and exit (orange) rates in period A by customer sector.
The sectors with the highest entry rates are K, P, and F; the sectors with the highest exit rates are K and P. 
For all customer sectors except NACE section K we observe net growth, with the highest growth rate for B, H and F.
On average, manufacturing sectors (\textless G) have lower turnover rates than service sectors (\textgreater G), with the exception of sector ``F - Construction".

\begin{figure*}[htb]
	\centering
	\includegraphics[width=0.95\textwidth]{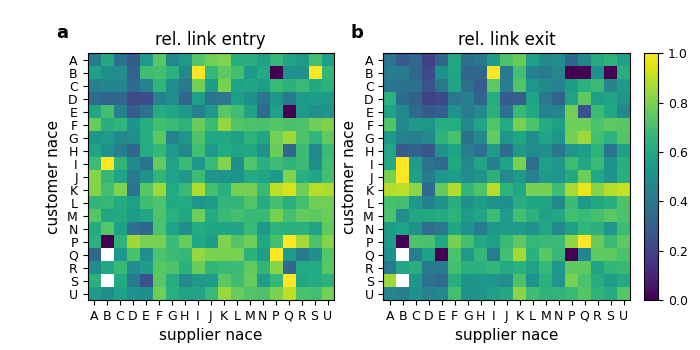}
	\caption{Link turnover in period A by supplier and customer industry combination.
		(a) Relative link entry rates by supplier (column) and customer (row) industry.
		(b) Relatie link exit rates by supplier (column) and customer (row) industry.
		The values are very heterogeneous even for single sectors.
	}
	\label{fig:SI_rel_entry_exit_sector_2d}
\end{figure*}

We further characterize the link turnover by calculating exit and entry rates for each supplier-customer sector combination. 
Figure~\ref{fig:SI_rel_entry_exit_sector_2d}a shows the relative link entry rates for period A by supplier and customer NACE section.
The entry rates are very heterogeneous, spanning values from $0$ to $1$. Very high or low values are typical for sector combinations with few observations.
The matrix is not symmetric and entry rates can be much higher in one direction than in the other, the mean absolute difference for reciprocal pairs is $0.056$.
In Fig.~\ref{fig:SI_rel_entry_exit_sector_2d}b we plot the relative link exit rates for period A by supplier and customer NACE section.
Again, the entry rates are very heterogeneous, spanning values from $0$ to $1$, with very high values typical for sector pairs with few links.
The matrix is not symmetric and exit rates can be much higher in one direction than in the other, the mean absolute difference for reciprocal pairs is $0.102$.
For both entry and exit the top left corner, denoting the turnover rates between manufacturing sectors (except F, i.e. \textless F), shows lower values than if service sectors are involved. 

\begin{figure*}[htb]
	\centering
	\includegraphics[width=0.8\textwidth]{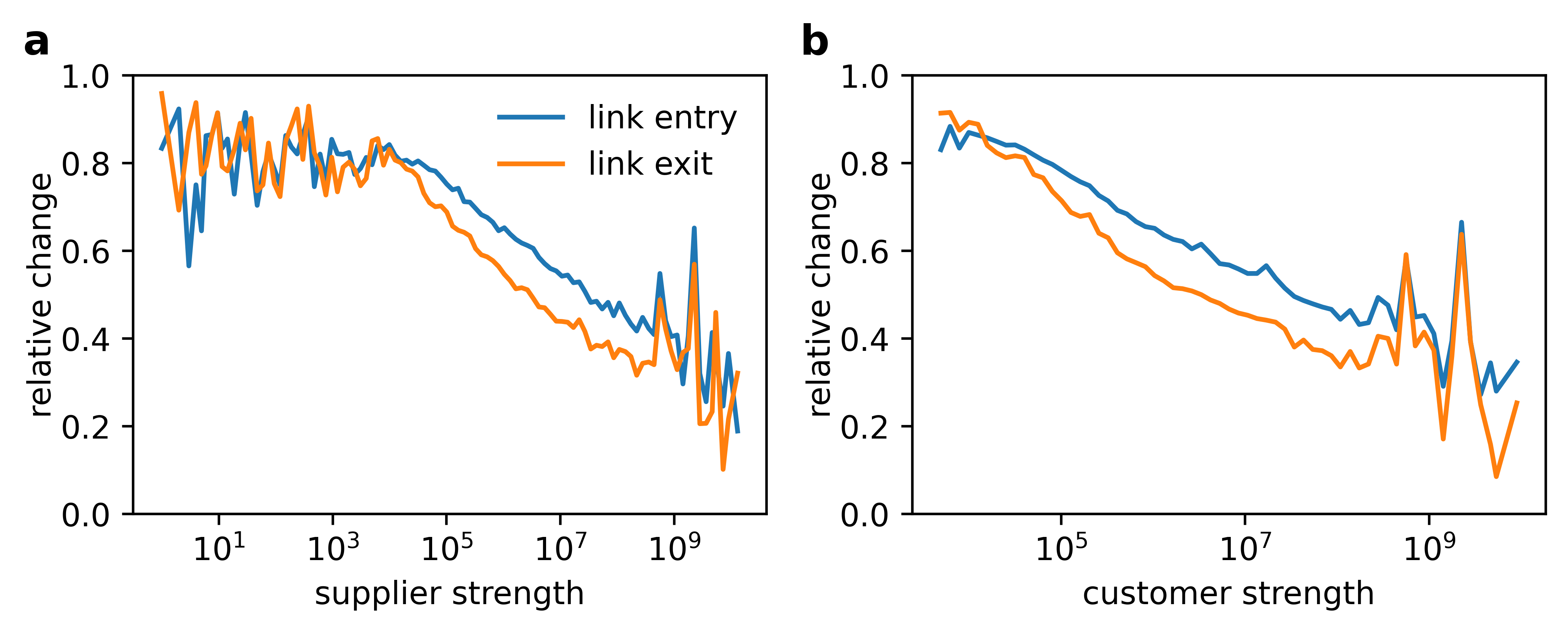}
	\caption{Link turnover in period A by supplier and customer strength.
		(a) Relative link entry (blue) and exit (orange) rate as function of the supplier strength, values are calculated using 100 logarithmic strength-bins.
		(b) Same as in (a) but as a function of customer sector. 
		In both panels we only plot bins that contain more than 100 links. Turnover drops after a threshold of around $10^5$kFT and is lower for large firms, both on the supplier and customer side.
	}
	\label{fig:SI_rel_entry_exit_strength_1d}
\end{figure*}

In Fig.~\ref{fig:SI_rel_entry_exit_strength_1d}a we plot relative link entry and exit rates as function of supplier strength, using logarithmic bins. Firms with low strength have generally very high customer turnover rates around 80\%, but above strength values around $10^5$kFT both relative entry and exit rates fall monotonically to around 30\% for the largest firms.
As the network is growing, entry rates are larger than exit rates for most strength values.
In Fig.~\ref{fig:SI_rel_entry_exit_strength_1d}b we show the same plot for customer strength. Firms with the lowest strength values have very high supplier turnover rates around 90\%, but both relative entry and exit rates fall monotonically to around 30\% for the largest firms.
As the network is growing, entry rates are larger than exit rates for most strength values.

\begin{figure*}[htb]
	\centering
	\includegraphics[width=0.95\textwidth]{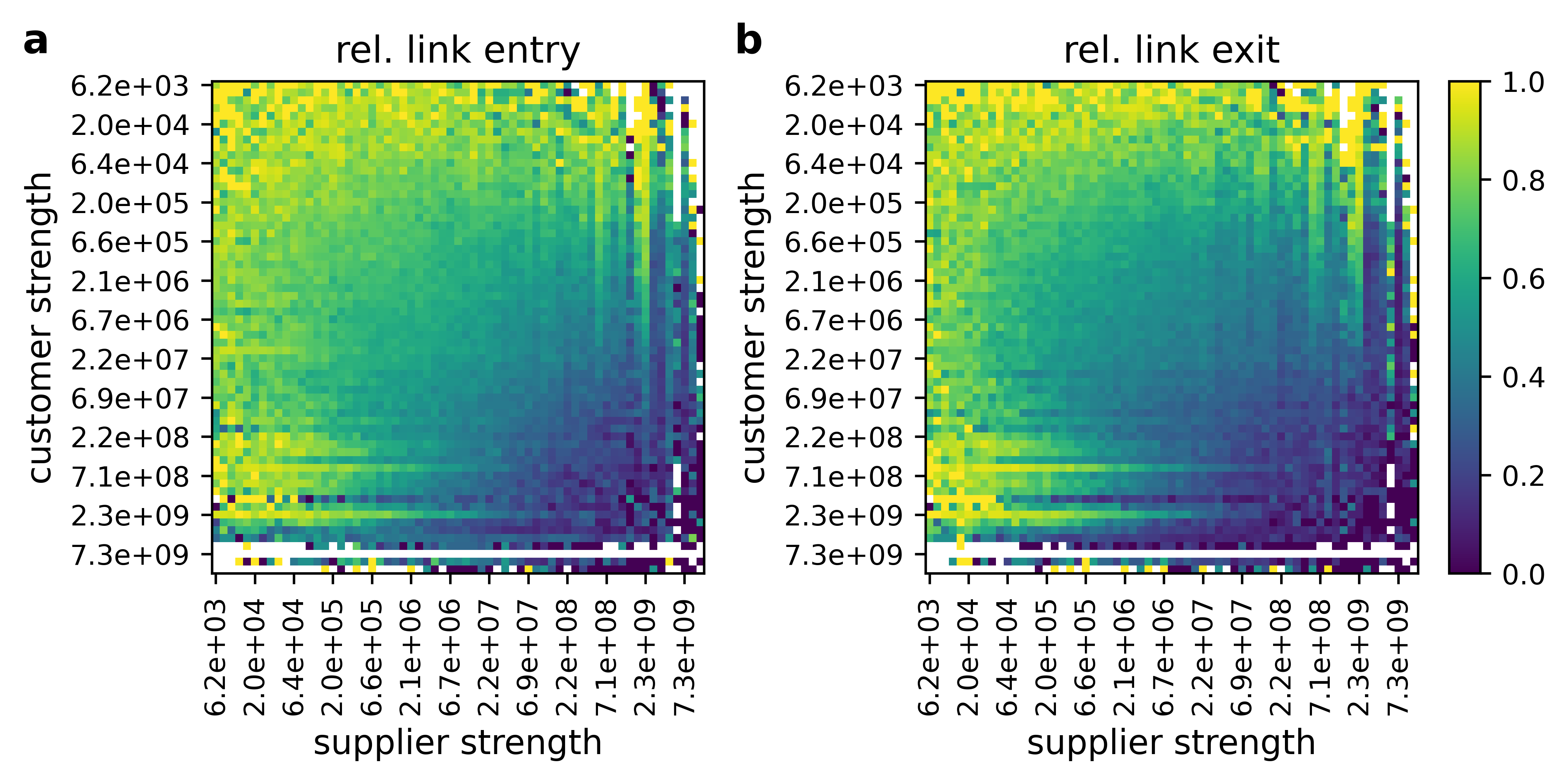}
	\caption{Link turnover in period A by supplier and customer strength, using logarithmic bins.
		(a) Relative link entry rates by supplier (column) and customer (row) strength.
		(b) Relatie link exit rates by supplier (column) and customer (row) strength.
		Turnover rates decay with higher strength of the involved parties. For low and high strength values there are few data points and the entry and exit rates are noisy.
	}
	\label{fig:SI_rel_entry_exit_strength_2d}
\end{figure*}

We further characterize the link turnover for each supplier-customer size combination. 
Figure~\ref{fig:SI_rel_entry_exit_strength_2d}a shows relative link entry rates for period A by supplier and customer strength bin (logarithmic bins).
The entry rates decay with higher strength of the involved parties. 
In Fig.~\ref{fig:SI_rel_entry_exit_strength_2d}b we show the same plot for relative link exit rates. Again, exit rates decay with higher strength of the involved parties. 
For both rates, there are few data points for low and high strength values and, hence, the entry and exit rates are noisy.

\begin{table}[ht]
	\centering
	\caption{Descriptions of NACE section codes.}
	\label{tab:SI_NACE_letters}
	
	\begin{tabular}{ll}
		\toprule
		{} &                                      Economic Area \\
		Code &                                                    \\
		\midrule
		A    &                  Agriculture, Forestry and Fishing \\
		B    &                               Mining and Quarrying \\
		C    &                                      Manufacturing \\
		D    &  Electricity, Gas, Steam and Air Conditioning S... \\
		E    &  Water Supply; Sewerage, Waste Management and R... \\
		F    &                                       Construction \\
		G    &  Wholesale and Retail Trade; Repair of Motor Ve... \\
		H    &                         Transportation and Storage \\
		I    &          Accommodation and Food Service Activities \\
		J    &                      Information and Communication \\
		K    &                 Financial and Insurance Activities \\
		L    &                             Real Estate Activities \\
		M    &  Professional, Scientific and Technical Activities \\
		N    &      Administrative and Support Service Activities \\
		O    &  Public Administration and Defence; Compulsory ... \\
		P    &                                          Education \\
		Q    &            Human Health and Social Work Activities \\
		R    &                 Arts, Entertainment and Recreation \\
		S    &                           Other Service Activities \\
		T    &  Activities of Households as Employers; Undiffe... \\
		U    &  Activities of Extraterritorial Organisations a... \\
		\bottomrule
	\end{tabular}
\end{table}

\subsection*{SI Text 2: Entry and exit rates}
We analyze monthly entry and exit rates for firms and links, respectively. The time a firm enters the network is denoted by its first overall occurrence and the time a firm exits by its last overall occurrence. A link exists if it is present in three or more months in a six month window. It enters on the first month of the first window where this condition is fulfilled and exits on the first month after the last window where this condition is met.

We fit a Poisson distribution with a seasonality adjusted rate, using the following procedure.
First, we count the number of monthly entry or exits, respectively, $X(t)$. 
Then we fit a linear trend to account for seasonality, $x(t) = \beta_1 \text{month}_t + \beta_0$, where $\text{month}_t$ denotes the number of the month, starting with 1 for January and ending with 12 for December.
The seasonality adjusted Poisson distribution is the mixture of twelve Poisson processes centered on $x(t)$, $$p(X) = \frac{1}{12}\sum_{\tau=1}^{12} P_{x(\tau)}(X) \quad \text{,}$$ where $P_{\lambda}(X)$ denotes the Poisson distribution with rate $\lambda$.

\begin{figure*}[htb]
	\centering
	\includegraphics[width=0.66\textwidth]{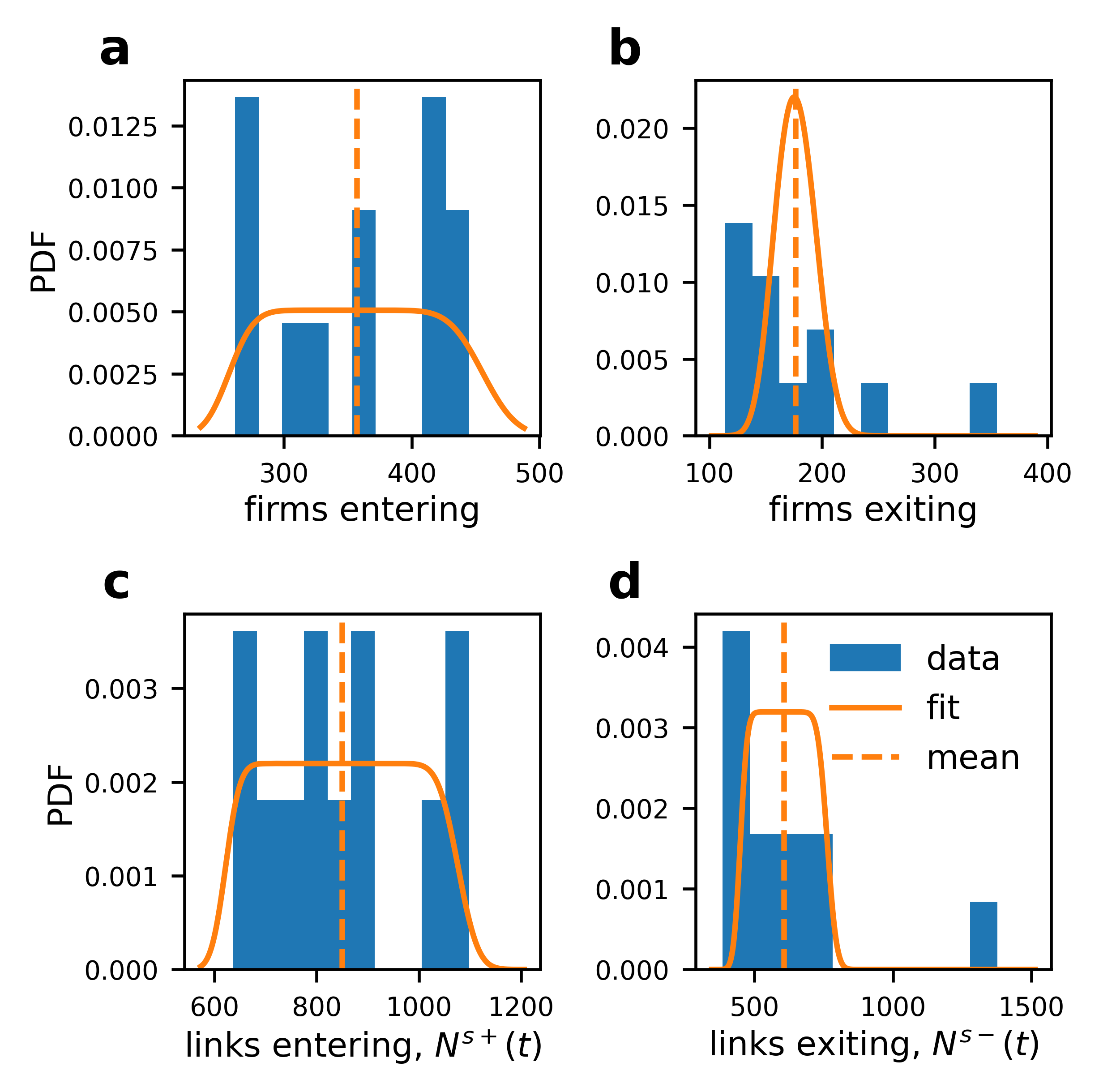}
	\caption{Monthly entry and exit rates in the Hungarian PNW. 
		(a) Empirical number of new firms per month (blue) in 2017, compared with a seasonality adjusted Poisson distribution (orange). The dashed vertical line denotes the empirical average new firms per time step, $348.9$.
		(b) Empirical number of firms exiting per month (blue) in 2017, compared with a seasonality adjusted Poisson distribution (orange). The dashed vertical line denotes the empirical average number of removed firms per time step, $167.6$.
		(c) Histogram of the empirical number of new links entering per month (blue) in 2017, compared with a seasonality adjusted Poisson distribution (orange). The dashed vertical line denotes the empirical average number of new links entering per time step, $867.8$.
		(d) Empirical number of links exiting per month (blue) in 2017, compared with a seasonality adjusted Poisson distribution (orange). The dashed vertical line denotes the empirical average number of removed links per time step, $606.9$.
	}
	\label{fig:entry_exit_rates}
\end{figure*}

In Fig.~\ref{fig:entry_exit_rates} we show the seasonality adjusted entry rates as solid orange lines. The entry processes in panels a and c are described well. For the exit processes in panels b and d, respectively, the symmetric seasonality adjusted Poisson distribution fails to capture the skew of the empirical distribution. In Tab.~\ref{tab:SI_entry_exit_rates} we report the average entry and exit rates for firms and nodes, respectively, separated by time period.

\begin{table}[ht]
	\centering
	\caption{Table with average entry/exit rates, separated by time period.}
	\label{tab:SI_entry_exit_rates}
	\begin{tabular}{l|rr|rr}
		         & \multicolumn{2}{c}{Period A} & \multicolumn{2}{c}{Period B} \\
		         &     entry & exit        &     entry &     exit \\
		\midrule
		links    &    867.8  & 606.9       &    1697.3 &   1301.5 \\
		nodes    &    348.9  & 167.6       &     605.1 &    360.2 \\
		\bottomrule
	\end{tabular}
\end{table}

\subsection*{SI Text 3: Details on link decay}

In this SI Text we provide details on the link decay process, in particular we discuss the functional form of the empirical link decay and we provide detailed results on sectoral decay rates for different time periods.

\begin{figure*}[htb]
	\centering
	\includegraphics[width=0.99\textwidth]{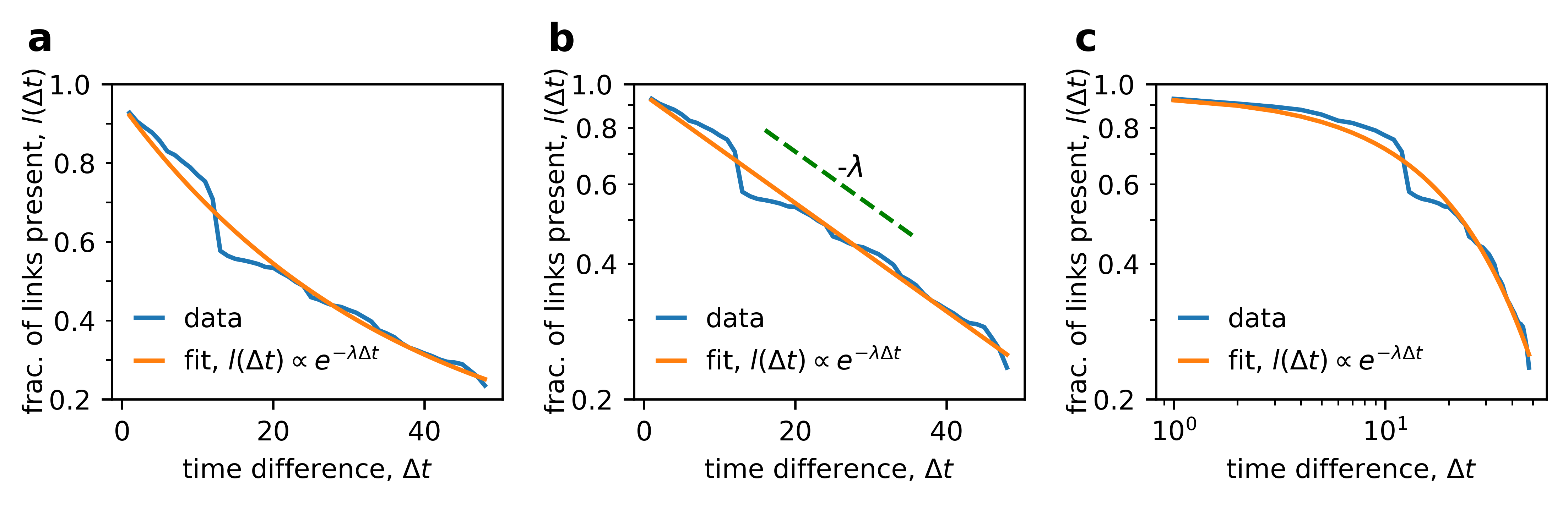}
	\caption{Relative fraction of links present $\Delta t$ months after Jan. 2017, $l(\Delta t)$, plotted using differently scaled axes. We plot both the data (blue) and the best fit (orange) using (a) linear,
		(b) semi logarithmic, and 
		(c) double logarthmic axes.
		If the data follows an exponential decay, it plots as a straight line in the semi-logarthmic plot, if it follows a power-law decay it plots as a straight line in the double logarithmic plot. The exponential fit describes the data best.
	}
	\label{fig:SI_link_exit_memory}
\end{figure*}

In the main text, we characterize link exit by counting the number of links present in the network at a time $t$, $L(t)$, and then counting again after a time $\Delta t$, $L(t+\Delta t)$. In Fig.~\ref{fig:SI_link_exit_memory}a we plot the relative number of links that remains $\Delta t$ months after $t_0 = \text{01/2017}$, $l(\Delta t) = L(t_0 + \Delta t)/L(t_0)$. The relative fraction of links after $\Delta t$ decays with a decreasing rate over time, but it is hard to tell the functional form of $l(\Delta t)$ from Fig.~\ref{fig:SI_link_exit_memory}a.

In the main text we chose an exponential fit to describe the link decay process. This corresponds to a memoryless process with constant decay rate that has the differential equation $l^\prime(t) = -cl$, which integrates to the exponential function $l(t) = k e^{-ct}$. Such a process manifests itself in a straight line in a semi-logarithmic plot, see Fig.~\ref{fig:SI_link_exit_memory}b, and describes the data well.

Alternatively, one could propose a process with memory, i.e. where the decay rate decays over time. The differential equation for such a process, $l^\prime(t) = -(c/t)l$ is solved by the power law $l(t) = k t^{-c}$, which would show up as a straight line in a double logarithmic plot. In Fig.~\ref{fig:SI_link_exit_memory}c we plot the the data (blue) on a double logarthimic axis and the exponential fit (orange). Both the fit and the data are clearly not a straight line, suggesting that a scaling law describing a process with memory is not a good description for links in the Hungarian PN.

In the main text we plot the decay rates for period A, In Tab.~\ref{tab:SI_decay_rates} we note the decay rates, $a_{\bar{w}_{0.5}}$, fitted at median weight, $\bar{w}_{0.5}$, for different time periods and NACE industries.

\begin{table}[ht]
	\centering
	\caption{Table with link decay rates, fitted at median weight.}
	\label{tab:SI_decay_rates}
	\begin{tabular}{l|rrr|rrr}
		\toprule
		{}   & \multicolumn{3}{c}{2017} & \multicolumn{3}{c}{2019} \\
		nace &      mean &        5\% &       95\% &      mean &        5\% &       95\% \\
		\midrule
		A    &  0.900995 &  0.885216 &  0.922171 &  0.900995 &  0.885216 &  0.922171 \\
		C    &  0.944346 &  0.940052 &  0.950722 &  0.944346 &  0.940052 &  0.950722 \\
		D    &  0.960553 &  0.956868 &  0.963245 &  0.960553 &  0.956868 &  0.963245 \\
		E    &  0.884572 &  0.799833 &  0.935946 &  0.884572 &  0.799833 &  0.935946 \\
		F    &  0.929888 &  0.917493 &  0.939161 &  0.929888 &  0.917493 &  0.939161 \\
		G    &  0.935416 &  0.933342 &  0.938008 &  0.935416 &  0.933342 &  0.938008 \\
		H    &  0.954000 &  0.951109 &  0.957627 &  0.954000 &  0.951109 &  0.957627 \\
		J    &  0.937026 &  0.930569 &  0.943492 &  0.937026 &  0.930569 &  0.943492 \\
		K    &  0.909709 &  0.865296 &  0.968903 &  0.909709 &  0.865296 &  0.968903 \\
		L    &  0.944099 &  0.941322 &  0.947060 &  0.944099 &  0.941322 &  0.947060 \\
		M    &  0.949814 &  0.946932 &  0.953335 &  0.949814 &  0.946932 &  0.953335 \\
		N    &  0.948567 &  0.944409 &  0.953664 &  0.948567 &  0.944409 &  0.953664 \\
		R    &  0.935168 &  0.914068 &  0.969111 &  0.935168 &  0.914068 &  0.969111 \\
		S    &  0.925820 &  0.925820 &  0.925820 &  0.925820 &  0.925820 &  0.925820 \\
		U    &  0.937560 &  0.934298 &  0.943567 &  0.937560 &  0.934298 &  0.943567 \\
		\bottomrule
	\end{tabular}
\end{table}

\subsection*{SI Text 4: Details on the average number of new suppliers per month}
- Add $\alpha$ at least in the baseline estimation for every time period!!

In the main text we fit a scaling relation $\langle N^{s+}\rangle = \alpha_0 k^{\alpha}$ for the average number of new suppliers per month, $\langle N^{s+}\rangle$.
For all periods the scaling exponent, $\alpha$, is not statistically significantly different from 1, {\bl XXX}.
We fix $\alpha = 1$ and show the fitted $\alpha-0$ for all NACE sections and time periods in Tab.~\ref{tab:SI_nsp_details}. We report the 90\% CI.

\begin{table}[ht]
	\centering
	\caption{Baseline supplier generation rate, $\alpha_0$, for NACE sections and time periods. We report the 90\% CI.}
	\label{tab:SI_nsp_details}
	\begin{tabular}{l|rrr|rrr}
		\toprule
		{}   & \multicolumn{3}{c}{2017} & \multicolumn{3}{c}{2019} \\
		nace & $\alpha_0$ &       5\% &      95\% & $\alpha_0$ &       5\% &      95\% \\
		\midrule
		A &  0.02042 &  0.01813 &  0.02300 &  0.02027 &  0.01817 &  0.02261 \\
		B &  0.09087 &  0.06955 &  0.11871 &  0.04608 &  0.03697 &  0.05744 \\
		C &  0.02631 &  0.02330 &  0.02971 &  0.02129 &  0.01906 &  0.02377 \\
		D &  0.01517 &  0.01306 &  0.01762 &  0.01758 &  0.01508 &  0.02049 \\
		E &  0.03454 &  0.02939 &  0.04060 &  0.03430 &  0.02930 &  0.04017 \\
		F &  0.02432 &  0.02115 &  0.02798 &  0.02552 &  0.02249 &  0.02896 \\
		G &  0.01151 &  0.01015 &  0.01305 &  0.00996 &  0.00888 &  0.01117 \\
		H &  0.01951 &  0.01703 &  0.02235 &  0.01808 &  0.01596 &  0.02049 \\
		I &  0.03672 &  0.03006 &  0.04486 &  0.03072 &  0.02590 &  0.03643 \\
		J &  0.02664 &  0.02328 &  0.03048 &  0.02532 &  0.02216 &  0.02892 \\
		K &  0.02911 &  0.02402 &  0.03529 &  0.03142 &  0.02489 &  0.03966 \\
		L &  0.01207 &  0.01019 &  0.01430 &  0.01350 &  0.01147 &  0.01589 \\
		M &  0.02564 &  0.02234 &  0.02942 &  0.02336 &  0.02048 &  0.02664 \\
		N &  0.02449 &  0.02143 &  0.02798 &  0.02162 &  0.01902 &  0.02458 \\
		O &  0.20000 &  0.04196 &  0.95334 &  0.16758 &  0.09007 &  0.31181 \\
		P &  0.06404 &  0.03514 &  0.11672 &  0.06021 &  0.03235 &  0.11204 \\
		Q &  0.21064 &  0.08507 &  0.52154 &  0.07282 &  0.03417 &  0.15520 \\
		R &  0.04672 &  0.03661 &  0.05963 &  0.05025 &  0.03964 &  0.06369 \\
		S &  0.05981 &  0.04642 &  0.07706 &  0.05668 &  0.04361 &  0.07367 \\
		U &  0.01434 &  0.01252 &  0.01643 &  0.01403 &  0.01233 &  0.01597 \\
		\bottomrule
	\end{tabular}
\end{table}

\subsection*{SI Text 5: Attachment kernel details}
In the main text we fit a scaling relation $A^k \propto k^{\beta}$ to determine the attachment kernel exponent, $\beta$.
In Tab.~\ref{tab:SI_Ak_details} we report $\beta$ for all NACE sections and time periods, with the respective limits of the 90\% CI.

\begin{table}[ht]
	\centering
	\caption{Attachment kernel scaling exponents, $\alpha$, by sector and time period. We report the respective limits of the 90\% CI.}
	\label{tab:SI_Ak_details}
	\begin{tabular}{l|rrr|rrr}
		\toprule
		{} & \multicolumn{3}{c}{2017} & \multicolumn{3}{c}{2019} \\
		{} & $\beta$ &         5\% &        95\% &  $\beta$ &         5\% &        95\% \\
		\midrule
		A &  1.638276 &    1.458471 &    1.818080 &  1.369997 &    1.257529 &    1.482465 \\
		B &  1.225872 &    0.829516 &    1.622228 &  0.860360 &    0.732422 &    0.988298 \\
		C &  0.916849 &    0.869831 &    0.963868 &  0.958597 &    0.927150 &    0.990044 \\
		D &  0.926334 &    0.852935 &    0.999734 &  0.823391 &    0.735633 &    0.911149 \\
		E &  1.004818 &    0.835834 &    1.173803 &  1.000170 &    0.845342 &    1.154999 \\
		F &  1.467235 &    1.348559 &    1.585911 &  1.484358 &    1.408171 &    1.560544 \\
		G &  1.263201 &    1.210515 &    1.315888 &  1.212597 &    1.176773 &    1.248420 \\
		H &  1.127075 &    1.035660 &    1.218490 &  1.071110 &    1.000068 &    1.142152 \\
		I &  1.555724 &    1.397347 &    1.714101 &  1.579479 &    1.362083 &    1.796875 \\
		J &  0.777848 &    0.682641 &    0.873056 &  0.843571 &    0.757417 &    0.929725 \\
		K &  0.807811 &    0.687261 &    0.928360 &  1.080271 &    0.979805 &    1.180737 \\
		L &  1.632160 &    1.493078 &    1.771243 &  1.468013 &    1.377034 &    1.558992 \\
		M &  1.490059 &    1.403764 &    1.576354 &  1.326500 &    1.245990 &    1.407011 \\
		N &  1.309293 &    1.232622 &    1.385964 &  1.308350 &    1.257149 &    1.359551 \\
		P &  1.574495 &    1.263537 &    1.885453 &  2.737200 &    1.841049 &    3.633350 \\
		R &  1.778426 &    1.437405 &    2.119447 &  1.563453 &    1.311219 &    1.815687 \\
		U &  0.948165 &    0.861630 &    1.034699 &  1.073029 &    1.027001 &    1.119057 \\
		S &  0.918622 &    0.455971 &    1.381273 &  1.295270 &    1.100198 &    1.490343 \\
		\bottomrule
	\end{tabular}
\end{table}

\subsection*{SI Text 6: Characterization of nodes that enter}
\begin{figure*}[htb]
	\centering
	\includegraphics[width=0.66\textwidth]{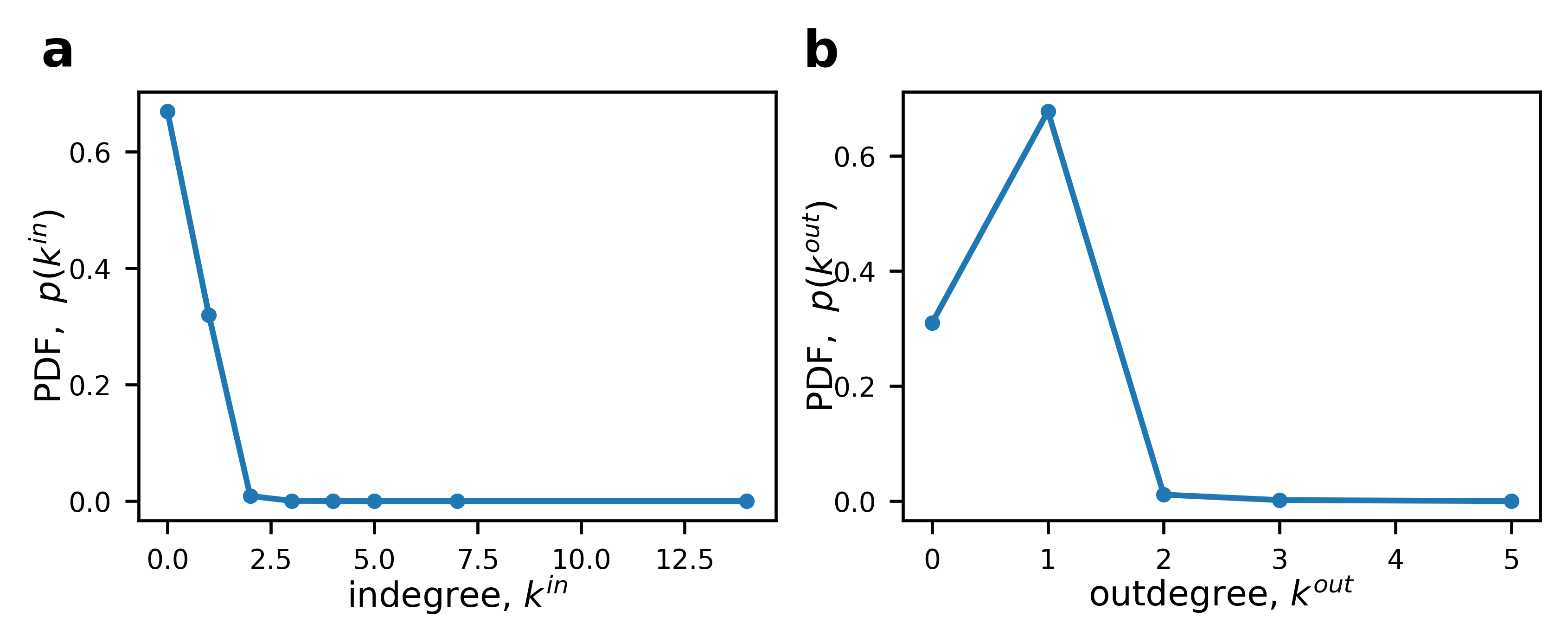}
	\includegraphics[width=0.66\textwidth]{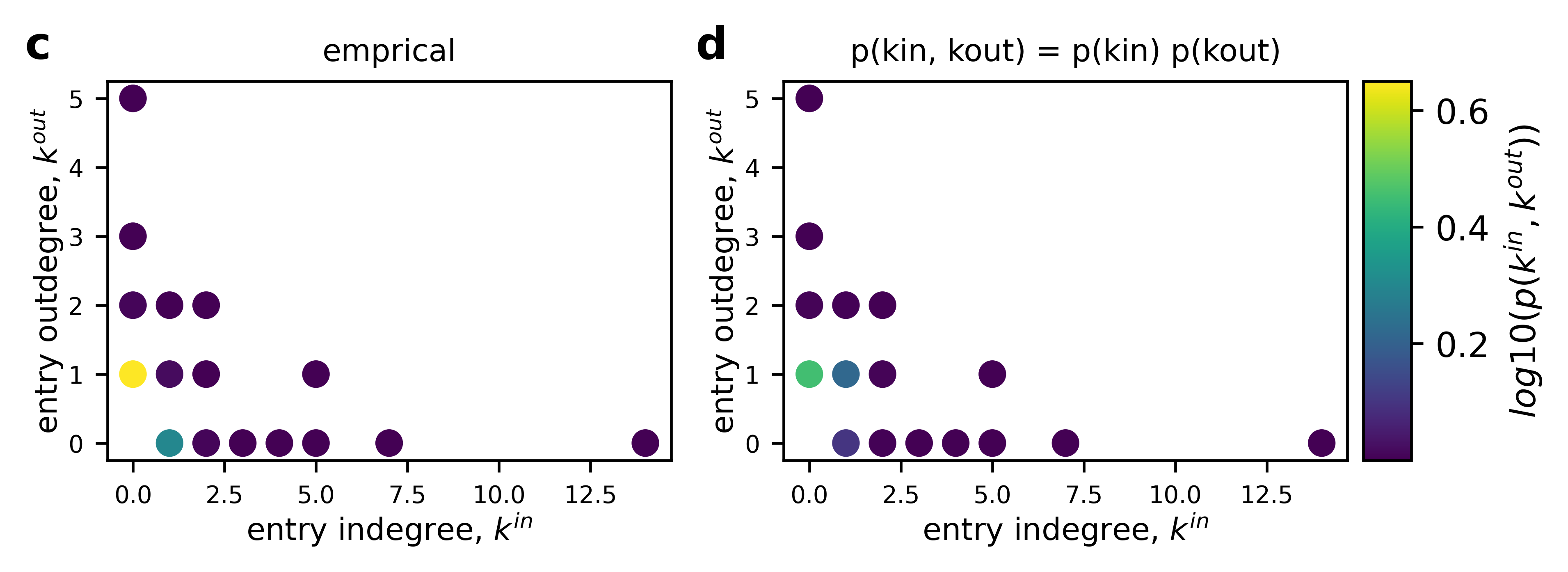}
	\caption{Degree upon entry.
		(a) Empirical marginal probability distribution of indegrees at node entry, $p(k^{in,0})$, and
		(b) empirical marginal probability distribution of outdegrees at node entry, $p(k^{out,0})$.
		(c) Empirical joint probability distribution of entry degrees, $p(k^{in,0}, k^{out,0})$.
		(d) Joint probability distribution calculated as product of the marginal distributions, $p^\prime(k^{in,0}, k^{out,0}) = p(k^{in,0})p(k^{out,0})$.
		The combination the degree combination $(k^{in,0}, k^{out,0}) = (1,1)$ is empirically underrepresented.
	}
	\label{fig:SI_degree_upon_entry}
\end{figure*}

When nodes enter, they are assigned a combination of in- and outdegree, $(k^{in,0}, k^{out,0})$, sampled from the empirical distribution of in- and outdegrees at node entry. 
In Fig.~\ref{fig:SI_degree_upon_entry}a we show the empirical probability distribution of indegrees at node entry, $p(k^{in,0})$. Most of the weight is concentrated on $k^{in,0}<2$, with the most common value at 0.
Figure~\ref{fig:SI_degree_upon_entry}b shows the empirical probability distribution of outdegrees at node entry, $p(k^{out,0})$. Again, most of the weight is concentrated on $k^{out,0}<2$, however, with the most common value at 1.
In the network generative model, we use the joint distribution, $p(k^{in,0}, k^{out,0})$, shown in Fig.~\ref{fig:SI_degree_upon_entry}c. Notably, not all values occur, and the entry events with highest $k^{in}$ and $k^{out}$, occur with no out- or in-links, respectively.
Most importantly, however, is the underrepresentation of the degree combination $(k^{in,0}, k^{out,0}) = (1,1)$, as can be seen by comparing the empirical distribution in panel Fig.~\ref{fig:SI_degree_upon_entry}c with the joint probability distribution calculated from the marginals under the assumption of independence, $p^\prime(k^{in,0}, k^{out,0}) = p(k^{in,0})p(k^{out,0})$, shown in Fig.~\ref{fig:SI_degree_upon_entry}d.
For our modelling exercise we truncate the empirical distribution at $k \le 3$, covering 99.8\% of the probability weight.

%
%

\subsection*{SI Text 7: Additional model results}
\subsubsection*{Additional results period A:}

\begin{figure*}[htb]
	\centering
	\includegraphics[width=0.99\textwidth]{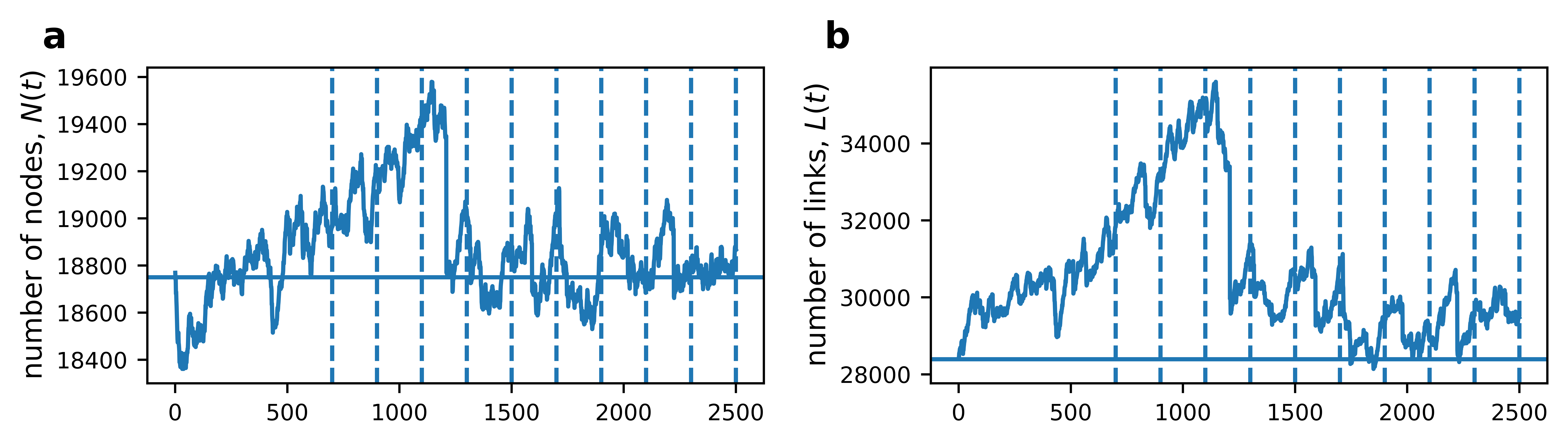}
	\caption{Evolution of the network size as function of model timestep.
		(a) Number of nodes, $N(t)$, as function of model time $t$. The horizontal line shows the average number of nodes in the empirical PN, the dashed vertical line marks the networks that were used as snapshots to study the network characteristics.
		(b) Number of links, $L(t)$, as function of model time $t$. The horizontal line shows the average number of links in the empirical PN, the dashed vertical lines mark the same snapshots as in (a).}
	\label{fig:SI_model_timeseries_NL_2017}
\end{figure*}

In this SI Text we provide additional model results for the year 2017, period A.
In Fig.~\ref{fig:SI_model_timeseries_NL_2017}a we show the number of nodes, $N(t)$, for every model timestep $t$. The model slightly underestimates the empirical number of nodes, $N_0$ (horizontal line).
Figure~\ref{fig:SI_model_timeseries_NL_2017}b shows the number of links, $L(t)$, for every model timestep $t$. The model fluctuates around the empirical number of links, $L_0$ (horizontal line).

\begin{figure*}[htb]
	\centering
	\includegraphics[width=0.66\textwidth]{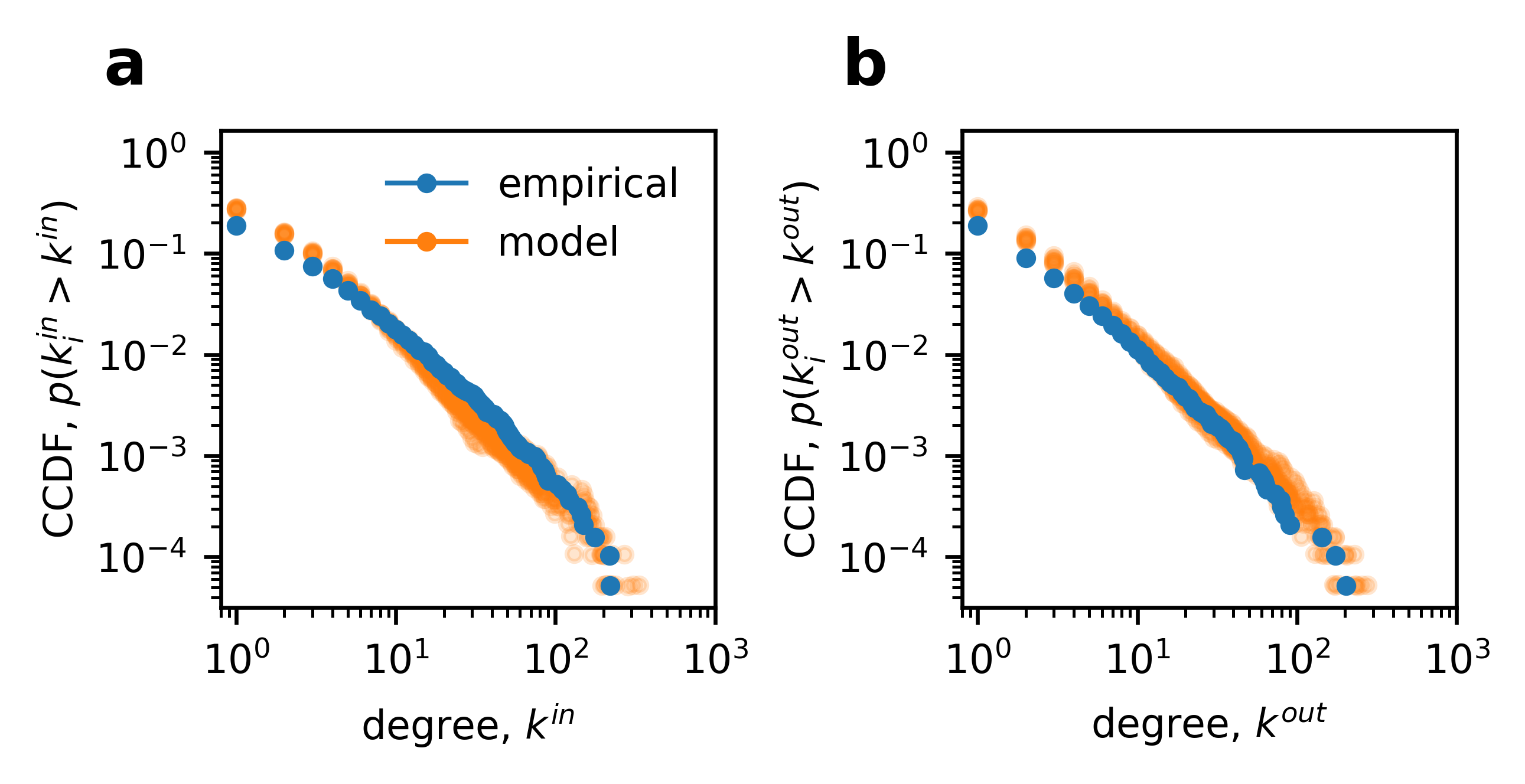}
	\caption{Modelled in- and outdegree for 2017. 
		(a) Counter cumulative distribution for $k^{in}$, $p(k^{in}_i > k^{in})$. The empirical distribution for Jan. 2017 is shown in blue, the distribution of ten model snapshots in orange.
		(b) Counter cumulative distribution for $k^{out}$, $p(k^{out}_i > k^{out})$. The empirical distribution for Jan. 2017 is shown in blue, the distribution of ten model snapshots in orange.}
	\label{fig:SI_indegree_outdegree_2017}
\end{figure*}

Figure~\ref{fig:SI_indegree_outdegree_2017}a compares the empirical indegree distribution (blue) with the indegree distribution of ten model snapshots (vertical lines in Fig.~\ref{fig:SI_model_timeseries_NL_2017}. The distributions match well, with a slight underestimation for high $k^{in}$.
Figure~\ref{fig:SI_indegree_outdegree_2017}b compares the empirical outdegree distribution (blue) with the outdegree distribution of ten model snapshots (vertical lines in Fig.~\ref{fig:SI_model_timeseries_NL_2017}. The distributions match well across the whole range of  $k^{out}$ values.

\begin{figure*}[htb]
	\centering
	\includegraphics[width=0.99\textwidth]{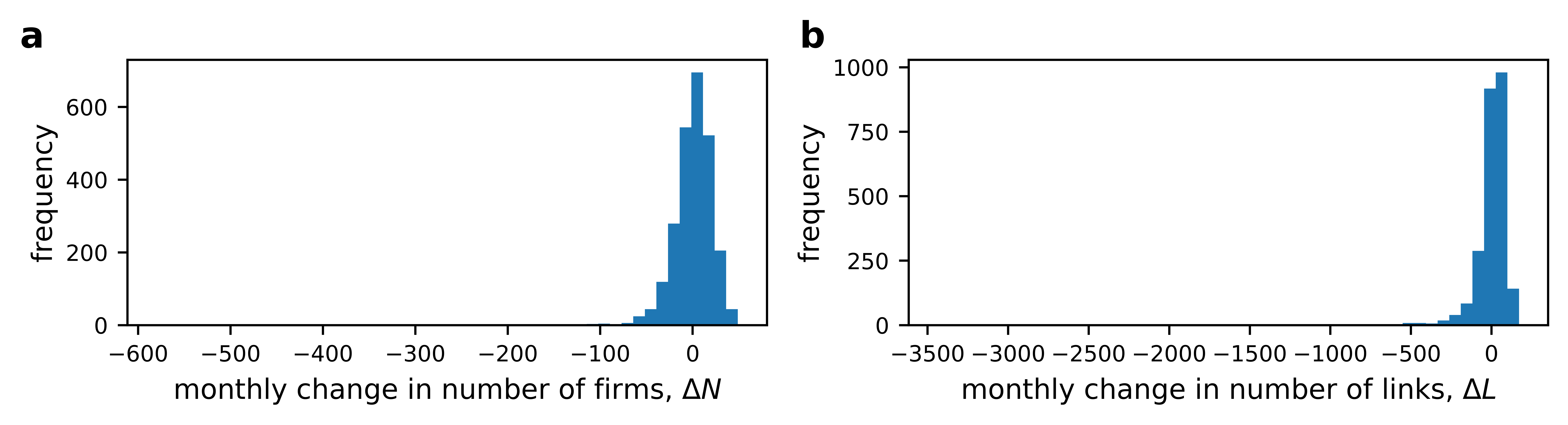}
	\caption{Distribution of monthly change in modeled network size variables in the model run shown in Fig.~\ref{fig:SI_model_timeseries_NL_2017} (period A).
		(a) Month-on-month difference in the number of firms, $\Delta N$, and
		(b) the number of links, $\Delta L$.
		Both quantities are heavily negatively skewed, with negative outliers caused by cascades of node exits. 
	}
	\label{fig:SI-model_DeltaNL}
\end{figure*}

In Fig.~\ref{fig:SI-model_DeltaNL} we plot the monthly changes in $N$ and $L$. Both quantities are negatively skewed with long negative tail. The skewness is caused by cascades of nodes exits, where a large node is removed and many of its neighbors become isolated and, hence, removed from the network.

\subsubsection*{Additional results period B:}

In this SI Text we provide the model results for period B, produced by the parameters specified in Tab.~\ref{tab:model_results}.

\begin{figure*}[htb]
	\centering
	\includegraphics[width=0.99\textwidth]{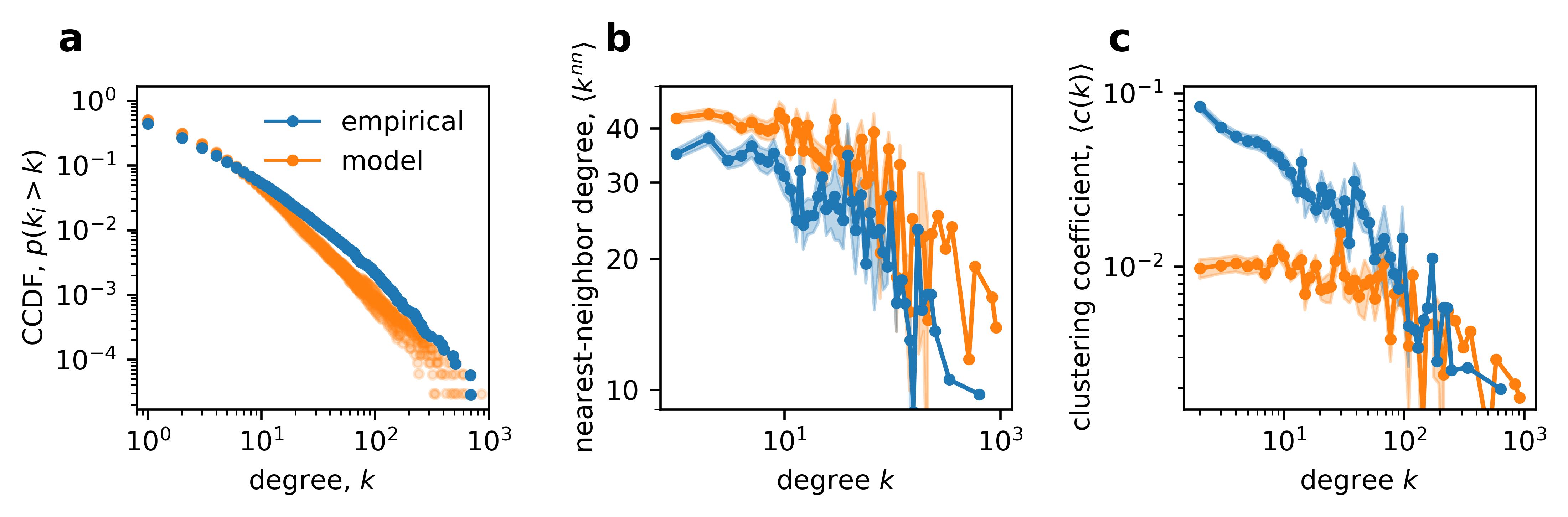}
	\caption{Model results 2019.
		(a) Counter cumulative degree distribution, $p(k_i \geq k)$ for snapshots generated by the supply network generating model (orange) and the empirical degree distribution in January 2019 (blue). The distributions are very similar.
		(b) Average nearest neighbor degree for the empirical (blue) and synthetic (orange) network calculated using linear degree bins for $k\le 10$ and logarithmic degree bins for $k > 10$, the shaded area denotes the standard error. Both networks are disassortative, however, the modeled nearest neighbor degree is higher than the empirical for all degree buckets.
		(c) Average local clustering coefficient for the eimpirical (blue) and synthetic (orange) network calculated using linear degree bins for $k\le 10$ and logarithmic degree bins for $k > 10$, the shaded area denotes the standard error. The local clustering coefficient is well captured for large degrees, but underestimated by up to 85\% for low degrees.
	}
	\label{fig:SI-model_results2019}
\end{figure*}

\begin{figure*}[htb]
	\centering
	\includegraphics[width=0.99\textwidth]{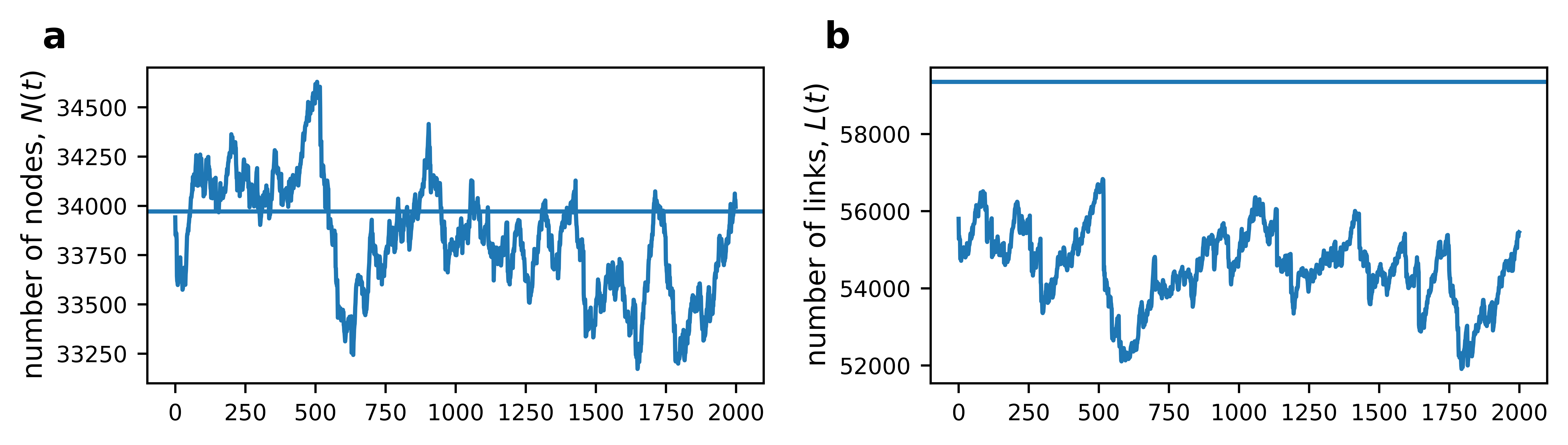}
	\caption{Evolution of the network size as function of model timestep for 2019.
		(a) Number of nodes, $N(t)$, as function of model time $t$. The horizontal line shows the average number of nodes in the empirical PN, the dashed vertical line marks the networks that were used as snapshots to study the network characteristics.
		(b) Number of links, $L(t)$, as function of model time $t$. The horizontal line shows the average number of links in the empirical PN, the dashed vertical lines mark the same snapshots as in (a).
		Both quantities slightly underestimate the empirical network size.}
	\label{fig:SI_model_timeseries_NL_2019}
\end{figure*}

\begin{figure*}[htb]
	\centering
	\includegraphics[width=0.66\textwidth]{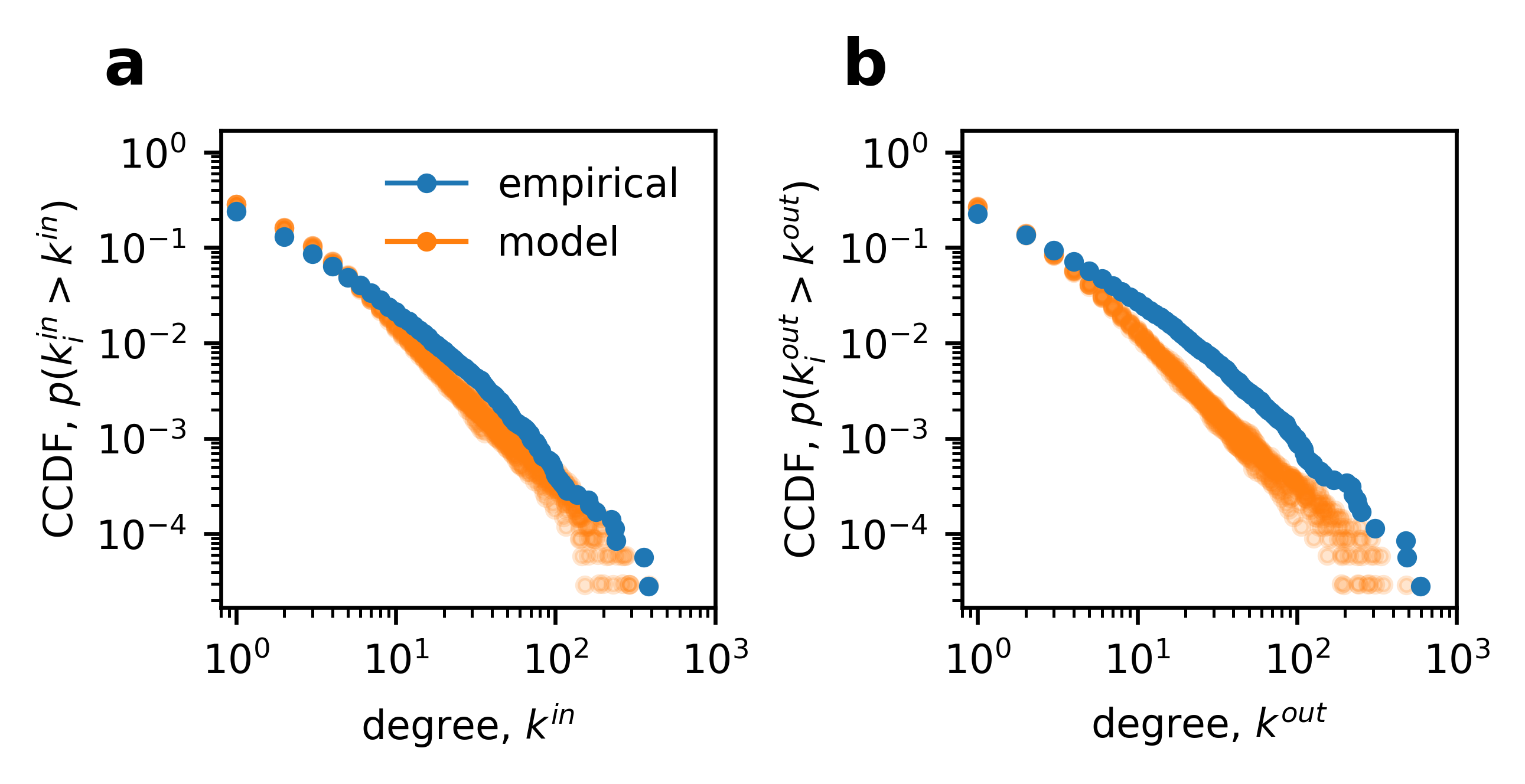}
	\caption{Modeled in- and outdegree for 2019. 
		(a) Counter cumulative distribution for $k^{in}$, $p(k^{in}_i > k^{in})$. The empirical distribution for Jan. 2019 is shown in blue, the distribution of ten model snapshots in orange.
		(b) Counter cumulative distribution for $k^{out}$, $p(k^{out}_i > k^{out})$. The empirical distribution for Jan. 2019 is shown in blue, the distribution of ten model snapshots in orange.}
	\label{fig:SI_indegree_outdegree_2019}
\end{figure*}

\subsubsection*{Additional results period C:}

In this SI Text we provide the model results for period C, produced by the parameters specified in Tab.~\ref{tab:model_results}.

\begin{figure*}[htb]
	\centering
	\includegraphics[width=0.99\textwidth]{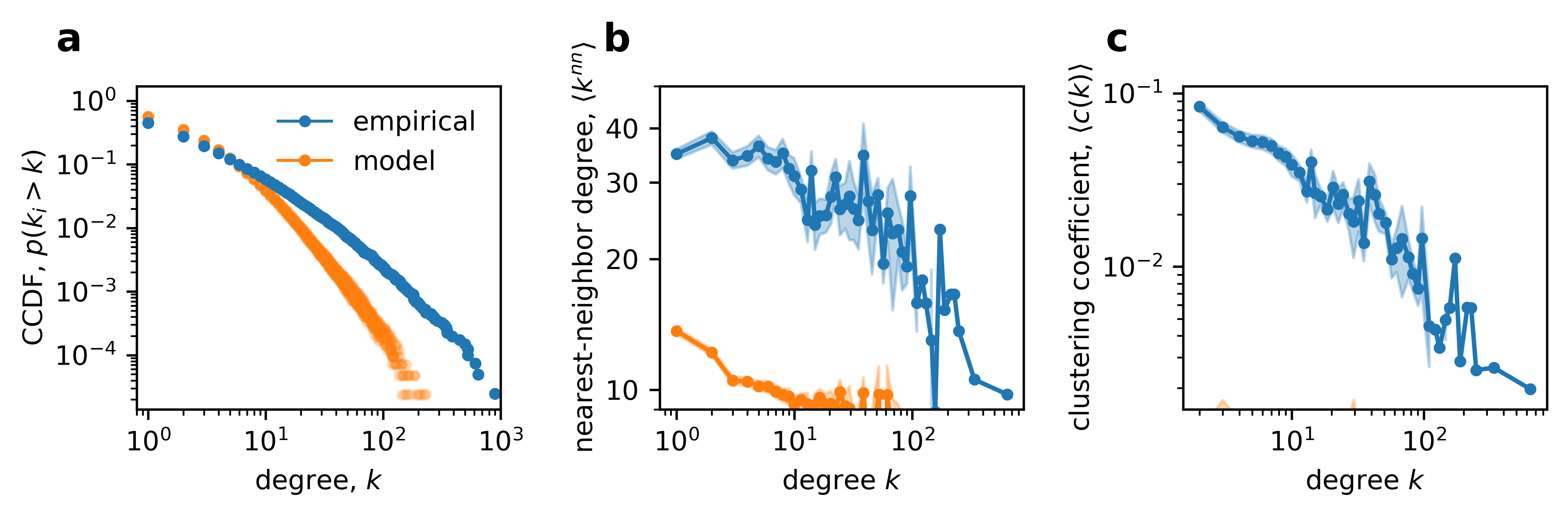}
	\caption{Model results 2021.
		(a) Counter cumulative degree distribution, $p(k_i \geq k)$ for snapshots generated by the supply network generating model (orange) and the empirical degree distribution in January 2021 (blue). The distributions are very similar.
		(b) Average nearest neighbor degree for the empirical (blue) and synthetic (orange) network calculated using linear degree bins for $k\le 10$ and logarithmic degree bins for $k > 10$, the shaded area denotes the standard error. Both networks are disassortative, however, the modeled nearest neighbor degree is higher than the empirical for all degree buckets.
		(c) Average local clustering coefficient for the eimpirical (blue) and synthetic (orange) network calculated using linear degree bins for $k\le 10$ and logarithmic degree bins for $k > 10$, the shaded area denotes the standard error. The local clustering coefficient is well captured for large degrees, but underestimated by up to 85\% for low degrees.
	}
	\label{fig:SI-model_results2021}
\end{figure*}

\begin{figure*}[htb]
	\centering
	\includegraphics[width=0.99\textwidth]{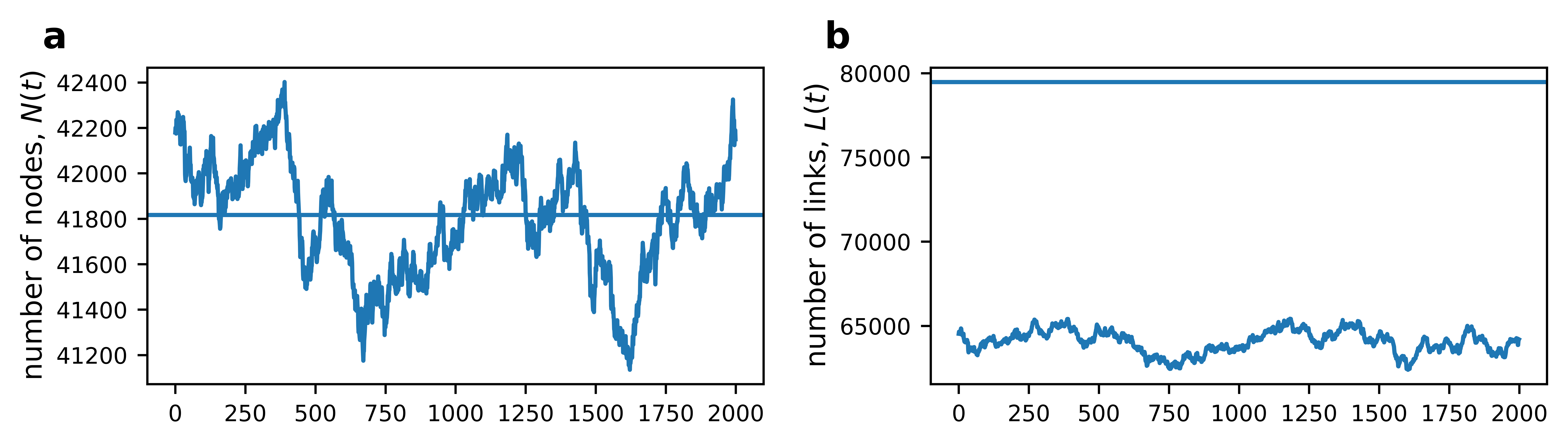}
	\caption{Evolution of the network size as function of model timestep for 2021.
	(a) Number of nodes, $N(t)$, as function of model time $t$. The horizontal line shows the average number of nodes in the empirical PN, the dashed vertical line marks the networks that were used as snapshots to study the network characteristics.
	(b) Number of links, $L(t)$, as function of model time $t$. The horizontal line shows the average number of links in the empirical PN, the dashed vertical lines mark the same snapshots as in (a).
	Both quantities slightly underestimate the empirical network size.}
	\label{fig:SI_model_timeseries_NL_2021}
\end{figure*}

\begin{figure*}[htb]
	\centering
	\includegraphics[width=0.66\textwidth]{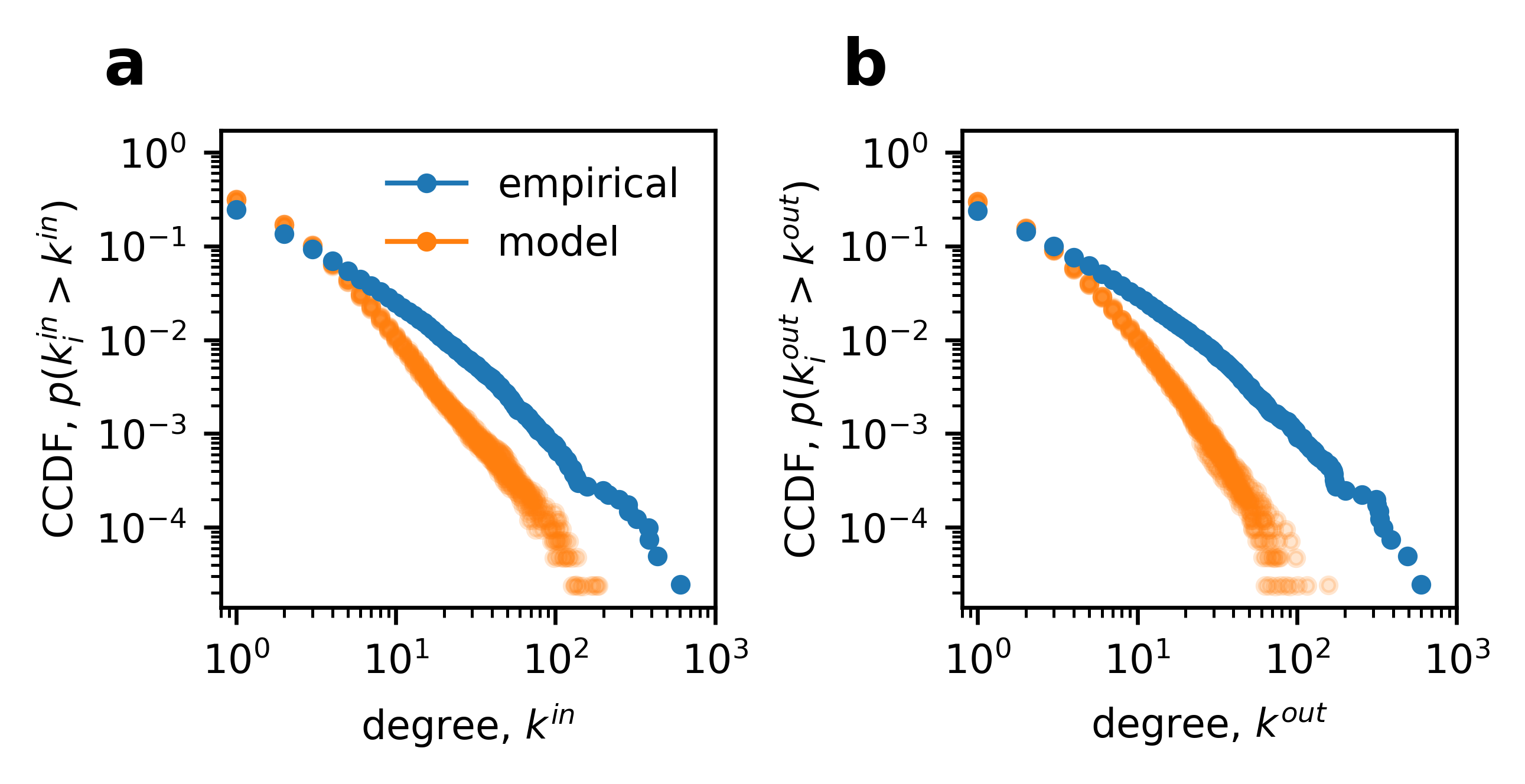}
	\caption{Modeled in- and outdegree for 2021. 
	(a) Counter cumulative distribution for $k^{in}$, $p(k^{in}_i > k^{in})$. The empirical distribution for Jan. 2021 is shown in blue, the distribution of ten model snapshots in orange.
	(b) Counter cumulative distribution for $k^{out}$, $p(k^{out}_i > k^{out})$. The empirical distribution for Jan. 2021 is shown in blue, the distribution of ten model snapshots in orange.}
	\label{fig:SI_indegree_outdegree_2021}
\end{figure*}

\subsection*{SI Text 8: Link filtering and the high systemic risk core}
The empirical ESRI profile presented in Fig.~\ref{fig:model_results_ESRI} is calculated using different specifications of the algorithm than in the original publication~\cite{diem2022quantifying}. In particular, the original algorithm corrects for the fact that the observed out-strength in the VAT network is lower than the firm's revenue because we don't observe final demand and exports and that the observed in-strength is lower than the firm's (input) consts, because we don't observe imports. Therefore, shocks spreading on the inter-firm network can not affect firms more than they are exposed to it. Further, we don't consider weights, because they are not modeled in the network generative model. Finally, the data used in \cite{diem2022quantifying} aggregates all transactions in a full year (with transactions in at least two distinct quarters), whereas we consider monthly data and a more restrictive filtering procedure.

\begin{figure*}[htb]
	\centering
	\includegraphics[width=0.99\textwidth]{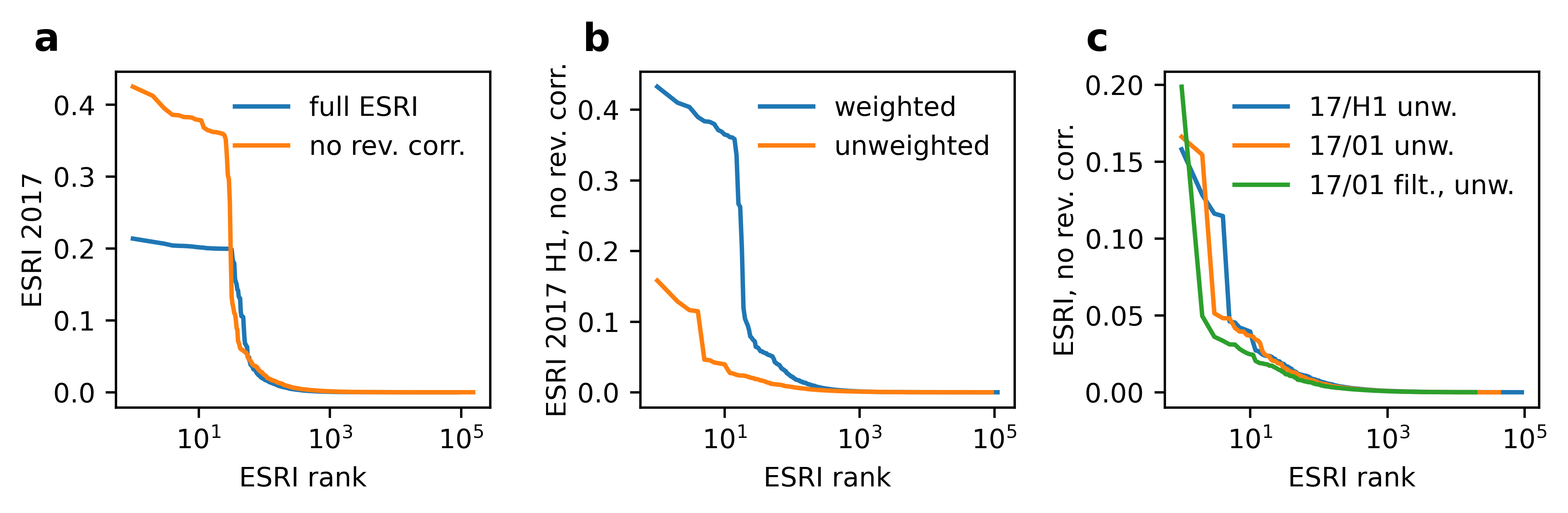}
	\caption{ESRI profile for different parametrizations and time periods.
		(a) Full ESRI profile for 2017 as in \cite{diem2022quantifying} (blue) and without revenue correction (orange).
		(b) ESRI for the first half year of 2017 without revenue correction for the weighted (blue) and unweighted (orange) network.
	    (c) ESRI without revenue correction and without weights for the first half year of 2017 (blue), for Jan. 2017 (orange) and for Jan. 2017 in the filtered network where links are only contained if they occur three times in a six month window (green).}
	\label{fig:SI_ESRI_no_plateau}
\end{figure*}

These changes result in the absence of a characteristic feature of ESRI, the formation of a \textit{high systemic risk core} of firms that all have a similar and high ESRI, visible as a plateau in the rank ordered distribution (the ESRI ``profile")~\cite{diem2022quantifying,reisch2022monitoring}. In Fig.~\ref{fig:SI_ESRI_no_plateau} we analyze the effects of changing the specifications on the ESRI profile.

In Fig.~\ref{fig:SI_ESRI_no_plateau}a we plot ESRI as calculated in \cite{diem2022quantifying} (blue) and ESRI calculated without revenue correction (orange). Firms are exposed to shocks in the PN much more strongly, rising the ESRI for the firms in the plateau from ca. $0.2$ to ca. $0.4$. The number of firms in the high systemic risk core, however, stays approximately the same.

We compare ESRI calculated on the weighted (blue) and unweighted (orange) network of the first half year of 2017 in Fig.~\ref{fig:SI_ESRI_no_plateau}b. The average ESRI, also for firms in the ESRI-plateu, is lowered dramatically, and the plateau contains fewer firms. Omitting link weights reduces the market share of many companies, resulting in a higher replaceability factor and lower ESRI.

Finally, in Fig.~\ref{fig:SI_ESRI_no_plateau}c we reduce the time window from the first half year in 2017 (blue), to January 2017 (orange) and January 2017 in the network with only stable links, i.e. links with at least three transactions in a six month window (green). The magnitue of ESRI does not change much, but the plateau gets shorter and vanishes for the filtered network. In every step we exclude more links, causing the systemic risk core to become disconnected. 

\subsection*{SI Text 9: Filtering procedure}
\begin{figure*}[htb]
	\centering
	\includegraphics[width=0.66\textwidth]{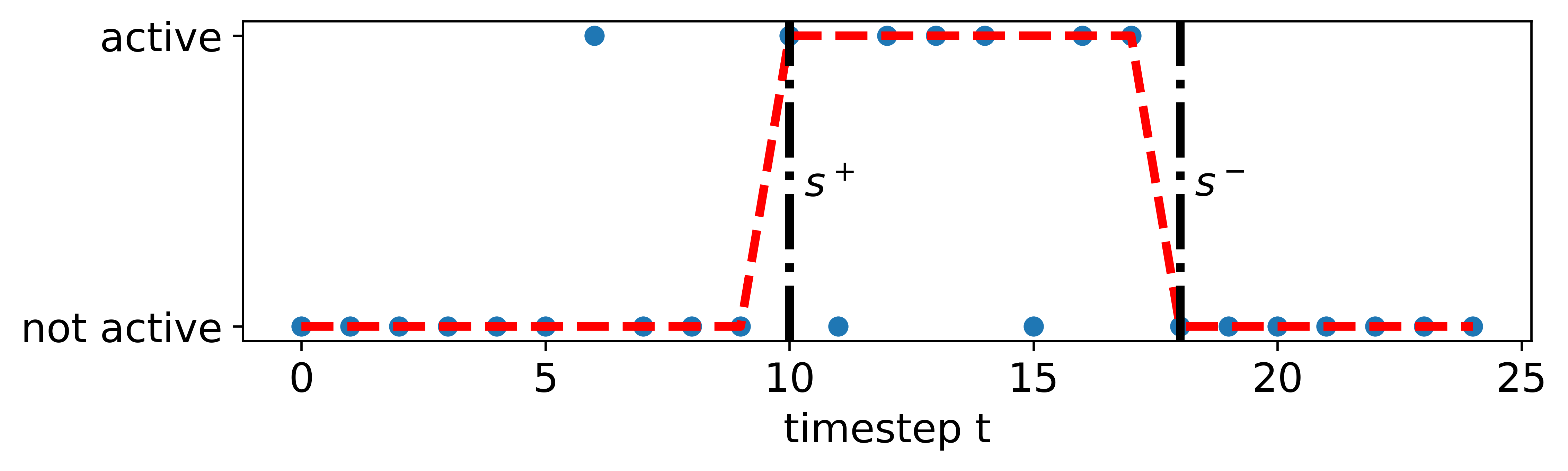}
	\caption{Filtering procedure. Blue dots show in which timestep (month) a transaction occurs. A link is considered active if it is activated at least three times in a six month window (red dashed line). We define link entry ($s^+$) as the first time step where the condition is fulfilled and exit ($s^-$) as the first time after the condition is fulfilled (black vertical lines).}
	\label{fig:SI_filtering_procedure}
\end{figure*}

To not be sensitive to one-off transactions or missed transactions in otherwise stable links, we filter for only stable links. In Fig.~\ref{fig:SI_filtering_procedure} we schematically illustratet he filtering procedure. The data consists of transaction data (blue dots) for each month. We define a link as active if it is present at least three times in a six month window (red dashed line). The link enters ($s^+$) at the first time step where the condition is fulfilled and exits ($s^-$) at the first time after the condition is fulfilled (black vertical lines). The filtering procedure is implemented using \texttt{R}'s \texttt{filter} function.

\end{document}